\documentclass[12pt]{article}
\usepackage[english]{babel}
\usepackage{csquotes}
\usepackage{amsthm}
\usepackage{amsfonts}
\usepackage{amsmath}
\usepackage{amssymb}
\usepackage{mathtools}
\usepackage{bbm}
\usepackage{graphicx}
\usepackage{wrapfig}
\usepackage{caption}
\usepackage{subcaption}
\usepackage{changepage}
\usepackage{adjustbox}
\usepackage{tcolorbox}
\usepackage{authblk}

\usepackage{natbib}
\usepackage[colorlinks=true,linkcolor=blue]{hyperref}
\hypersetup{urlcolor=blue, citecolor=blue, colorlinks=true, linkcolor=blue} 

\usepackage{pgfplots, pgfplotstable}
\pgfplotsset{width=5cm, compat=1.3}
\pgfplotstableread[col sep = comma]{BrownianBridges_High.csv}\brownianBridgeDataHighCorr
\pgfplotstableread[col sep = comma]{BrownianBridges_Low.csv}\brownianBridgeDataLowCorr
\pgfplotstableread[col sep = comma]{BrownianBridges_Zero.csv}\brownianBridgeDataZeroCorr
\pgfplotstableread[col sep = comma]{BrownianBridges_Negative.csv}\brownianBridgeDataNegativeCorr

\usepackage{tikz}
\usetikzlibrary{shapes, decorations}

\theoremstyle{plain}
\newtheorem{theorem}{Theorem}
\newtheorem{proposition}{Proposition}

\theoremstyle{definition}
\newtheorem{definition}{Definition}

\theoremstyle{remark}
\newtheorem{remark}{Remark}

\newcommand{\norm}[1]{\left\lVert #1 \right\rVert}
\newcommand{\round}[1]{\left( #1 \right)}
\renewcommand{\square}[1]{\left[ #1\right]}
\newcommand{\curly}[1]{\left\{ #1 \right\}}
\newcommand{\abs}[1]{\left| #1 \right|}

\newcommand{\Norm}[2]{\mathcal{N}\round{#1,\,#2}}

\newcommand{\Exp}[1]{\mathbb{E}\round{#1}}

\newcommand{\Cov}[2]{\mathbb{C}\text{ov}\round{#1,#2}}
\newcommand{\Corr}[2]{\mathbb{C}\text{or}\round{#1,#2}}

\newcommand{\N}[0]{\mathbb{N}}
\newcommand{\Z}[0]{\mathbb{Z}}
\newcommand{\R}[0]{\mathbb{R}}
\newcommand{\ul}[1]{\underline{#1}}
\newcommand{\ol}[1]{\overline{#1}}

\DeclareMathOperator{\diag}{diag}
\DeclareMathOperator{\Var}{\mathbb{V}ar}
\DeclareMathOperator{\ls}{ls}
\DeclareMathOperator{\lf}{lf}

\begin{document}

\title{Temporally-Evolving Generalised Networks and their Reproducing Kernels}
\author[1]{Tobia Filosi}
\author[1]{Claudio Agostinelli}
\author[2,3]{Emilio Porcu}
\affil[1]{Department of Mathematics, University of Trento, Italy}
\affil[2]{Department of Mathematics, Khalifa University, Abu Dhabi, UAE}
\affil[3]{School of Computer Science and Statistics, Trinity College Dublin, Ireland}

\maketitle

\begin{abstract}
This paper considers generalised network, intended as networks where (a) the edges connecting the nodes are nonlinear, and (b) stochastic processes are continuously indexed over both vertices and edges. Such topological structures are normally represented through special classes of graphs, termed graphs with Euclidean edges. We build generalised networks in which topology changes over time instants. That is, vertices and edges can disappear at subsequent time instants and edges may change in shape and length. We consider both cases of linear or circular time. For the second case, the generalised network exhibits a periodic structure. Our findings allow to illustrate pros and cons of each setting. Generalised networks become semi-metric spaces whenever equipped with a proper semi-metric. Our approach allows to build proper semi-metrics for the temporally-evolving topological structures of the networks. Our final effort is then devoted to guiding the reader through appropriate choice of classes of functions that allow to build proper reproducing kernels when composed with the temporally-evolving semi-metrics topological structures.
    
\end{abstract}

\section{Introduction}
    \subsection{Context}
        \label{ssec:introContext}
        Data complexity is certainly one of the main aspects to address in the Data Science revolution framework. In particular, we call {\em 3D} complexities those aspects related to Data structure, Data dimension and Data domain. The present paper concerns the latter of these aspects and delves into the problem of graphs whose nodes and edges evolve dynamically over time. \par
        Data analysis on graphs became ubiquitous in both Statistics and Machine Learning communities. For the first, recent contributions as in \cite{anderes2020}, \cite{moradi}, \cite{BADDELEY2021100435} and \cite{BADDELEY} witness the importance of graph structures for georeferenced data, being realisations of either geostatistical or point processes. As for the machine learning community, the amount of literature is huge. Open Graph Benchmark (OGB) is a comprehensive set of challenging and realistic benchmark datasets with the aim to facilitate scalable, robust and reproducible graph machine learning (ML) research \citep{hu2020open}. An overview of ML methodological approaches on graphs is provided by \cite{chami2022machine}. Topological complexities under the framework of data analytics are discussed in \cite{stankovic2020data}. Excellent surveys about ML on graphs are oriented to angles as different as large scale challenges \citep{hu2021ogb}, automated ML \citep{zhang2021automated}, representation learning \citep{hamilton2017representation}, relational ML for knowledge graphs \citep{nickel2015review} and higher order learning \citep{agarwal2006higher}, to mention just a few. \par
        In the great majority of contributions, the process is assumed to be defined exclusively over the vertices of the graph. The extension to processes that are continuously defined over both graphs and edges requires substantial mathematical work. \par
        This paper focuses on graphs with Euclidean edges \citep{anderes2020}, being an ingenious topological structure that allows to generalise linear networks to nonlinear edges. Further, the process defined over such structures can have realisations over any point over the edges, and not only in the nodes. Roughly, these are graphs where each edge is associated with an abstract set in bijective correspondence with a segment of the real line. This provides each edge with a Cartesian coordinate system to measure distances between any two points on that edge. \par
        There has been increasing attention on generalised networks in ML \citep{alsheikh2014machine,  georgopoulos2014distributed, hamilton2017representation, pinder2021gaussian, borovitskiy2022isotropic}, spatial data \citep{cressie2006spatial, gardner2003predicting, ver2006spatial, peterson2013modelling, peterson2007geostatistical, montembeault2012impact} and  point processes \citep{xiao2017modeling, perry2013point, deng2014ginibre, baddeley2017stationary}. For all these contributions, \emph{time} is not part of the game. \par    
        Reproducing kernel Hilbert space (RKHS) methods \citep{hofmann2008kernel} had a considerable success in both ML and Statistics community, and the reader is referred to \cite{hofmann2006review}, \cite{kung2014kernel} and to \cite{pillonetto2014kernel} for excellent overviews. \par
        RKHS methods require \emph{kernels}, being positive semidefinite functions defined over a suitable input space. A customary assumption for such kernels is that of isotropy, \emph{i.e.} the kernel depends only on the distance between any pair of points belonging to the input space. There is a rich literature at hand for the case of the input space being a $d$-dimensional Euclidean space \citep[see the celebrated work of][]{schoenberg}, and the reader is referred to the recent review by \cite{porcu2023mat}. Non Euclidean domains have a more recent literature, and we mention \cite{PBG16} as well as \cite{borovitskiy_isotropic_2023} and \cite{borovitskiy_matern_2020} for recent contributions. For such cases, the \emph{distance} between the points is no longer the Euclidean distance, but the geodesic distance. \par
        The \emph{tour de force} by \cite{anderes2020} has allowed to define isotropic reproducing kernels for generalised network, by working with two metrics: the geodesic and the resistance metric. Elegant isometric embedding arguments therein allow to provide sufficient conditions for given classes of functions to generate a legitimate reproducing kernel through composition with either of the two metrics. See the discussion in Section \ref{sec:MathematicalBackground}.    
    
    \subsection{Graphs cross time and temporally-evolving graphs}
        \label{ssec:graphsTime}
        Data on generalised networks are usually repeatedly observed over time. For such a case, it is customary to consider the input space, $X$, as a product (semi) metric space, with two separate metrics: the geodesic (or the resistance) metric for the graph, and the temporal separation for time. This approach has considerable advantages as it simplifies the mathematical architecture considerably. Unfortunately, under such a setting, the graph topology is invariant over time. This fact implies that nodes cannot disappear (nor new nodes can appear at arbitrary future instant times), and the shape and length of the edges do not evolve over time. Reproducing kernels for such a case have been recently proposed by \cite{porcu2022nonseparable} and by \cite{tang}. When the input space is a product space equipped with separate metrics, the kernel is component-wise isotropic when it depends, on the one hand, on a suitable distance over the graph and, on the other hand, on temporal separation. For details, the reader is referred to \cite{porcu2022nonseparable}. 
        For the case of static metric graphs, we mention the impressive approach in \cite{bolin}.
        \par 
        Outside the reproducing kernel framework, scientists have been mainly focused on the topological structure of temporally dynamical networks \citep{hanneke_discrete_2010}.   This fact boosted for a wealth of related approaches, ranging from community detection methods \citep{mankad_structural_2013, cherifi_community_2019} to link prediction \citep{lim_link_2019, divakaran_temporal_2020} to structural changes detection \citep{rossi_modeling_2013}. The common feature of the above contributions is that they consider the stochastic evolution of the structure (nodes and edges) of the graphs as the basis for the inference, whilst they do not usually allow for \emph{processes} defined over those graphs. \par
    
    \subsection{Linear or circular time? Might graphs be periodic?}
        \label{ssec:introCircular}
        This paper consider time-evolving graphs. While the choice of linear time does not need any argument---this is what most of the literature does---this paper argues that periodically-evolving graphs have a reason to exist. Our first argument to advocate in favour of a periodic construction is that it suits perfectly to several real-world phenomena, where both linear-time evolution (\textit{e.g.} long-term trends) and cyclic oscillations (\textit{e.g.} seasonal components) might happen. To make an example, consider temperatures in a given geographical area: apparently there might be strong correlations between: (i) contiguous spatial points at a given time, which are represented by means of spatial edges; (ii) the same points considered at contiguous times, which are represented by means of temporal edges between temporal layers and (iii) the same points considered at the same periods of the year, which are considered in the model as they are exactly the same point in the temporally evolving graph. 
        In many real-world applications, the network underlying a system is only partially observable. As a consequence, it could be hard or impossible to specify the whole time-evolving (not periodic) network in cases of long time series. The periodic assumption, when all in all reasonable, may be a great help in this circumstance as well.
    
    \subsection{Our contribution}
        \label{ssec:contribution}
        This paper provides the following contributions.
        \begin{enumerate}
            \item We provide a mathematical construction for graphs with Euclidean edges having a topology that evolves over time. That is, the number of vertices can change over time, as well as the shape and length of the edges. Remarkably, our construction allows for stochastic processes that are continuously indexed over both vertices and edges.
            \item We start by considering the case of linear time. After providing the structure for a temporally evolving graph, we devote substantial mathematical effort to build a suitable semi-metric over it. This is achieved at the expense of a sophisticated construction through a Gaussian bridge that is interpolated through the edges in the spatio-temporal domain.
            \item As an implication of the previous points, we obtain suitable second order properties (hence, the reproducing kernel) associated with such a process.
            \item The previous steps are then repeated for a periodic time-evolving graph.
            \item We prove that the construction with linear time might have a counterintuive property: adding temporal layers can change the inter-space distances between points in the graph. This might be a problem in terms of statistical inference, as carefully explained through the paper. We show that the periodic construction does not present such an inconvenience.
            \item Our findings culminate by guiding the reader through handy constructions for kernels defined over these graphs.
        \end{enumerate}
        
        We should mention that our contribution differentiates with respect to earlier literature in several direction. In particular: 
        \begin{enumerate}
            \item We allow the topology of the graph to evolve over time (whatever linear or circular). Previous contributions where graphs with Euclidean edges are considered use either a static graph \citep{anderes2020, bolin2022gaussian} or a graph having a topology that is invariant with respect to time \citep{porcu2022nonseparable, tang}. Hence, we provide a very flexible framework in comparison with earlier literature.
            \item We allow stochastic processes to be continuously defined over the graph. This is a substantial innovation with respect to a massive literature from both Statistics and ML, where the graph topology can vary according to some probability law associated with the nodes, but not to processes defined over the nodes. 
        \end{enumerate}
            
        The structure of the paper is the following. Section \ref{sec:MathematicalBackground} recalls the main mathematical objects that will be used. Section \ref{sec:TEGraphsWithEE} builds the skeleton of our construction, \emph{i.e.} time-evolving graphs, which are exploited in Sections \ref{sec:resMetricLinearTime} and \ref{sec:resMetricPeriodicTime}, where time-evolving graphs with Euclidean edges are defined for the linear time and circular time cases, respectively. Section \ref{sec:reproducingKernels} illustrates how it is possible to build kernels on such a structure and present some examples. Finally, Section \ref{sec:conclusion} concludes the paper. \par
        In addition, in Appendix \ref{app:mathBackground} we recall some mathematical definitions used throughout. In Appendix \ref{app:isotropicKernelsGeneral}, we recall and then extend some significant results by \cite{anderes2020} about the definition of kernels on arbitrary domains that are used in this manuscript, and present them under a general and easy-to-handle perspective. As proofs are rather technical, we defer them to Appendix \ref{app:proofs} for a neater exposition of the main text.

\section{Mathematical background}
    \label{sec:MathematicalBackground}
    This material is largely expository and provides the necessary mathematical background needed to understand the concept illustrated in the main text. For the unfamiliar reader, Appendix \ref{app:mathBackground} provides basic definitions and concepts used in network theory.
    \subsection{Gaussian random fields over semi-metric spaces}
        Let us begin with a brief introduction about Gaussian random fields \citep{stein-book} \par
        Let $X$ be a non-empty set and let $k:X\times X\to \R$. Then $k$ is a \emph{positive semi-definite} function (or a \emph{kernel}, or a \emph{covariance function}) if and only if, for all $n\in\N^+$, $x_1,\dots,x_n\in X$ and $a_1,\dots,a_n\in\R$, 
        \begin{equation}
            \label{eq:positiveSemidefiniteKernel}
            \sum_{i=1}^n\sum_{j=1}^n a_i a_j k(x_i,x_j)\geq 0.
        \end{equation}
        If, in addition, whenever the above relation is an equality, then necessarily $a_1=\dots =a_n = 0$, $k$ is \emph{(strictly) positive definite}. \par
        For  $X$ as above, we denote $Z$ a real-valued random field, \textit{videlicet}: for each $x \in X$, $Z(x)$ is a real-valued random variable. Then $Z$ is called \emph{Gaussian} if, for all $n\in \N^+$ and $x_1,\dots,x_n\in X$, the random vector $\boldsymbol{Z}:=(Z(x_1),\ldots,Z(x_n))^{\top}$, with $\top$ denoting the transpose operator,  follows a $n$-variate Gaussian distribution.  \par
        A Gaussian random field $Z$ on $X$ is completely determined by its first two moments: the mean function
        \begin{align*}
            \mu_Z: X &\to \R \\
            x & \mapsto \Exp{Z(x)},
        \end{align*}
        with $\mathbb{E}$ denoting stochastic expectation, and the covariance function (kernel)
        \begin{align*}
            k_Z: X \times X &\to \R \\
            (x_1,x_2) &\mapsto \Cov{Z(x_1)}{Z(x_2)}.
        \end{align*}
        A necessary and sufficient condition for a function $k_Z$ to be a covariance function (a kernel) of some random field $Z$ is to be positive semi-definite. \par
        For $X$ as above, we define a mapping $d:X\times X\to \R$. Then $(X,d)$ is called \emph{semi-metric space} (or, equivalently, $d$ is called a \emph{semi-metric} on $X$) if the following conditions hold for each $x,y\in X$: 
        \begin{enumerate}
            \item $d(x,y)\geq 0$,
            \item $d(x,y)=0\Longleftrightarrow x=y$,
            \item $d(x,y)=d(y,x)$.
        \end{enumerate}
        In addition, $(X,d)$ is called a \emph{metric space} (or, equivalently, $d$ is called a \emph{metric} on $X$) if it is a semi-metric space and the triangle inequality holds, namely, for all $x,y,z \in X$:
        \[
            d(x,y)+d(y,z)\geq d(x,z).
        \]
        The covariance function $k_Z$ is called isotropic for the semi-metric space $(X,d)$ if there exists a mapping $\psi: D_X^{d} \to \mathbb{R}$ such that $k_Z(x,y)=\psi(d(x,y))$, for $x,y \in X$. Here, $D_X^d:=\curly{d(x_1,x_2)\,:\,x_1,x_2\in X}$ is the diameter of $X$. See Appendix \ref{app:isotropicKernelsGeneral} on how to construct isotropic kernels on arbitrary domains. For a Gaussian random field $Z$ on $X$, we define its \emph{variogram} $\gamma_Z: X \times X \to \R$ through 
        \begin{equation}
            \label{eq:Variogram}
            \gamma_Z(u_1,u_2):=\Var\round{Z(u_1)-Z(u_2)}, \qquad u_1,u_2 \in X,
        \end{equation}
        with $\Var$ denoting \emph{variance}. The celebrated work of \cite{schoenberg} proves that $\gamma_Z$ is a variogram if and only if the mapping $\exp(-\gamma_Z(\cdot,\cdot))$ is positive definite on $X \times X$. \par
        Let $(X_1,d_1)$ and $(X_2,d_2)$ be two semi-metric spaces. Then, the triple $(X_1 \times X_2, d_1, d_2)$ is called a \emph{product semi-metric space}. \cite{menegatto_gneiting_2020} define isotropy over a product semi-metric space through continuous functions $\psi:D_{X_1}^{d_1} \times D_{X_2}^{d_2} \to \R$ such that, for $(x_1,x_2), (x_1', x_2') \in X_1 \times X_2$,
        \begin{equation} 
            \label{eq:productKernel}
            ((x_1,x_2), (x_1', x_2')) \mapsto \psi(d_1(x_1,x_1'), d_2(x_2,x_2')),
        \end{equation}
        is positive definite. \par
        The above definition naturally arises from spatio-temporal settings: suppose we have a \emph{static} semi-metric space $(X,d)$ that represents some spatial structure and $(T,d_T)$ representing time, where $T \subseteq \R$ and, usually, $d_T(t_1,t_2)=\abs{t_1-t_2}$. In such a case, Equation (\ref{eq:productKernel}) can be re-adapted to define kernels. This is the setting adopted by \cite{porcu2022nonseparable} and by \cite{tang}.
        
        \begin{remark}
        We deviate from earlier literature and we instead consider a metric space $X_t$ that evolves over time, $t \in T$. Hence, our domain is written as
            \begin{equation*}
                \curly{(x_t,t)\,:\, x_t \in X_t,\, t \in T},
            \end{equation*}
            where $t$ describes {\em time}, and the graph coordinate $x_t$ is constrained on the space $X_t$. Such a framework entails a way more sophisticated construction to equip such a space with a proper metric.
        \end{remark}

    \subsection{Graphs with Euclidean edges}
        \label{ssec:GraphsEuclEdges}
        We start with a formal definition of graphs with Euclidean edges. We slightly deviate from the definition provided by \cite{anderes2020}, for the reasons that will be clarified subsequently. For a definition of graph, see \ref{def:simpleConnGraph} in the Appendix.
        \begin{definition}[Graph with Euclidean edges]
            \label{def:graphWithEE}
            Consider a simple, connected and weighted graph $G=(V, E, w)$, where $w:E\to \R^+$ represents the weight mapping. Then, $G$ is called a \emph{graph with Euclidean edges} provided that the following conditions hold.
            \begin{enumerate}
                \item Edge sets: Each edge $e\in E$ is associated to the compact segment (also denoted by $e$) $[0,\ell(e)]$, where $\ell(e):=w(e)^{-1}$ may be interpreted as the \emph{length} of the edge $e$. 
                \item Linear edge coordinates: Each point $u\in e=(\ul u,\ol u)$ is uniquely determined by the endpoints $\ul u$ and $\ol u$ of $e$ and its relative distance $\delta_e(u):=\frac{u}{\ell(e)}=u\,w(e)$ from $\ul u$, that is $u=\round{\ul u, \ol u, \delta_e(u)}$, so that $\ul u=(\ul u,\ol u,0)=(\ol u,\ul u,1)$ and $\ol u=(\ul u,\ol u,1)=(\ol u,\ul u,0)$.
            \end{enumerate}
        \end{definition}
        Henceforth, we shall assume the existence of a total order relation on the set of vertices $V$ and that every edge is represented through the ordered pair $(v_1,v_2)$, where $v_1<v_2$. In particular, for each $u\in e$, the endpoints of $e$, $\ul u$ and $\ol u$ satisfy the relation $\ul u<\ol u$. \par
        A relevant fact is that our setting deviates from \cite{anderes2020}. In particular, our Definition \ref{def:graphWithEE}  does not require any \emph{distance consistency} opposed to \citet[Definition 1, (d)]{anderes2020}. The reason is that our setting does not need a bridge between geodesic and resistance metrics. A second relevant fact is that we have restricted the space of possible bijections from each edge onto closed intervals with orientation. We restrict to linear bijections: the main reason is that the focus of this paper is not to explore isometric embeddings, but to provide suitable topological structures evolving over time, and attach to them stochastic processes. \par
        \begin{figure}[tp]
            \centering
            \includegraphics[width = \textwidth]{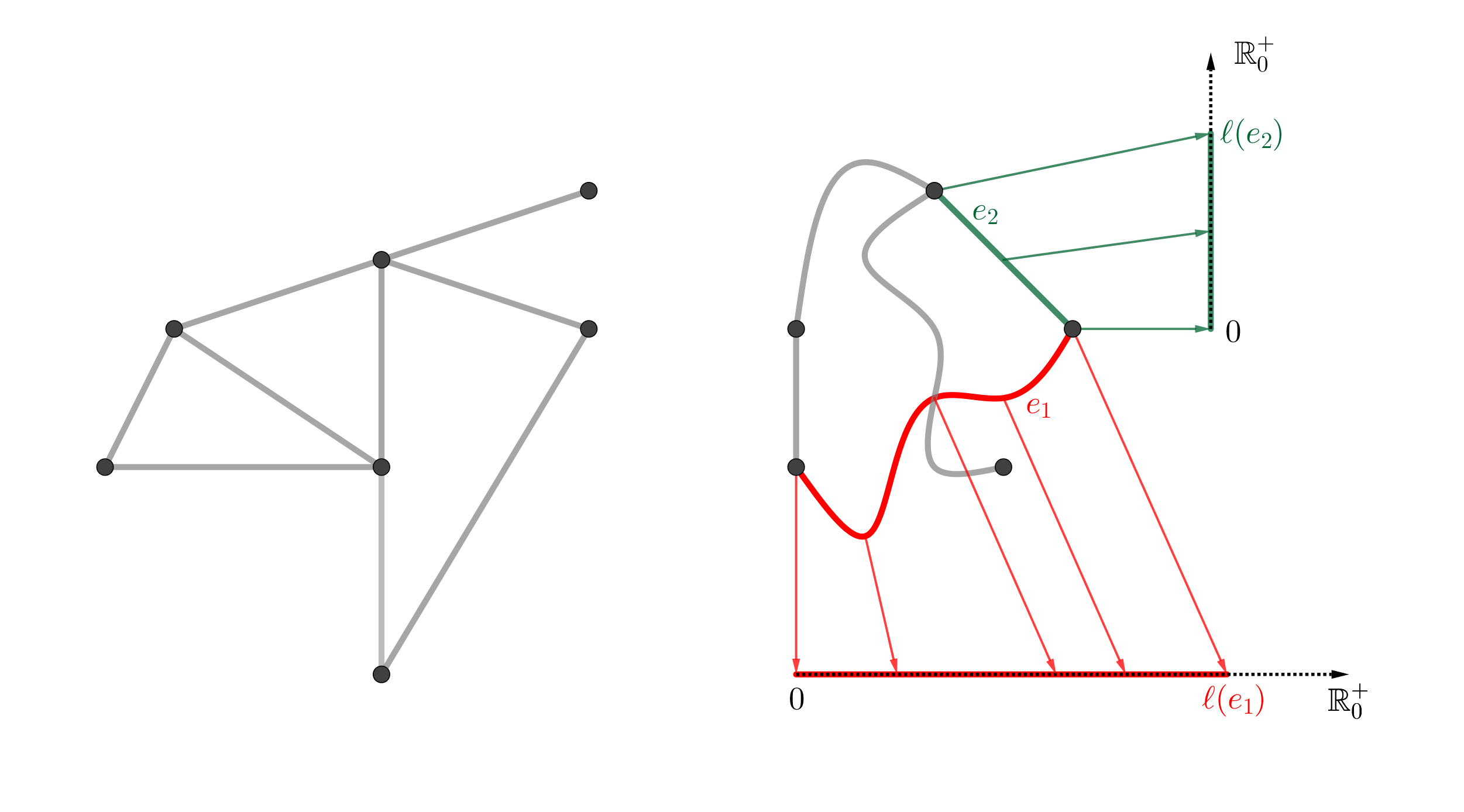}
            \caption{Left: a linear network. Right: a graph with Euclidean edges, where the bijections between the edges $e_1$ and $e_2$  and their respective real segments $[0,\ell(e_1)]$ and $[0,\ell(e_2)]$ are stressed.}
            \label{fig:linearNet_graphEE}
        \end{figure}
        As a final remark, we stress that the framework introduced through Definition \ref{def:graphWithEE} is way more general than linear networks, for at least two reasons: (a) in our framework, the weights of each edge be chosen independently from the others, and (b) our framework needs no restriction on the network structure, \textit{e.g.}, as shown in Figure \ref{fig:linearNet_graphEE} (right), edges may cross without sharing the crossing point.

    \subsection{Graph laplacian and resistance metric}
        The resistance metric has been widely used in graph analysis, as it is more natural than the shortest-path metric when considering flows or transport networks, where multiple roads between two given points may share the total flow. In order to define the classic effective resistance distance for an undirected and connected graph, we briefly report its definition and its mathematical construction. \par
        Let $G=(V,E,w)$ be a simple, weighted and connected graph (see Definition \ref{def:simpleConnGraph} in the Appendix) and let $W$ its adjacency matrix, that is: $W(v_1,v_2)=w((v_1,v_2))$, where we set $w((v_1,v_2))=0$ whenever $v_1\not\sim v_2$. In addition, for each node $v\in V$, we define its \emph{degree} as the sum of the weights of the edges adjacent to it. Let $D$ be the degree matrix of $G$, i.e. the diagonal matrix where each diagonal element is the degree of the corresponding vertex. Then, the \emph{laplacian matrix} (or simply \emph{laplacian}) of $G$ is the matrix $L:=D-W$, namely the matrix $L:V\times V\to \R$ having entries 
        \begin{equation}
            L(v_1,v_2)=\begin{cases}
                -w((v_1,v_2))\qquad &\text{if $v_1\ne v_2$}\\
                \sum_{u\in V} w((v_1, u)) \qquad &\text{if $v_1=v_2$}.
            \end{cases}
        \end{equation} \par
        Laplacian matrices enjoy several properties (see, for instance, \cite{devriendt_effective_2022}): they are symmetric, diagonally dominant, positive semidefinite and singular with exactly one null eigenvalue, corresponding to the eigenvector $\boldsymbol{1}_{n}$. Furthermore, they have non-positive off-diagonal entries and positive main-diagonal entries.\par 
        A graph $G=(V,E)$ is called a \emph{resistor graph} if the edges $e \in E$ represent electrical resistors and the nodes represent contact points. Given a resistor graph, the \emph{effective resistance distance} $R$ between two vertices is defined as the voltage drop between them when injecting one Ampere of current in one and withdrawing one Ampere from the other. \par
        Several mathematical formulations of this concept have been provided, and the reader is reminded, among many others, to \citet[Subsection 2.1]{jorgensen_hilbert_2010}. Throughout, we follow \cite{ghosh_minimizing_2008}. Let $G=(V,E)$ be a resistor graph. For each $v_1\sim v_2\in V$, let $r(v_1,v_2)\in\R^+$ denote the resistance of the resistor that connects $v_1$ and $v_2$. In addition, for each $v_1,v_2\in V$, define the weight (which plays the role of the physical conduttance)
        \[
            w((v_1,v_2)):=
            \begin{cases}
                \frac{1}{r((v_1,v_2))}\qquad &\text{if }v_1\sim v_2\\
                0\qquad &\text{if } v_1 \not \sim v_2.
            \end{cases}
        \]
        Let $L$ be the laplacian matrix of $G$ with the above-defined weights, $L^+$ its Moore-Penrose generalised inverse (see Definition \ref{def:MoorePenroseInverse} in the Appendix). Finally let $e_{v_i}$ denote the vector with all zeroes, except a one at position $v_i$. Then the effective resistance distance $R$ between two nodes $v_1$ and $v_2$ enjoys the following expression:
        \begin{equation}
            \label{eq:resistanceDistanceDef}
            R(v_1,v_2)=(e_{v_1}-e_{v_2})^\top L^+ (e_{v_1}-e_{v_2}).
        \end{equation}

\section{Time-evolving graphs with Euclidean edges}
    \label{sec:TEGraphsWithEE}
    Defining a time-evolving graph with Euclidean edges requires some mathematical formalism. While keeping such a formalism below, we shall then provide some narrative in concert with some graphical representation to have an intuition of how these graphs work. \par
    Even though the underlying rationale of our construction is quite natural and intuitive, the mathematical description of such an object is quite involved as it requires several of steps and a substantial formalism. Hence, we provide a sketch of our procedure to help the reader in the Box below. 

    \begin{tcolorbox}[title = \bf{A sketch of our construction}]
        \begin{enumerate}
            \item \label{item:sketch1}Define a time-evolving graph as a properly defined sequence of graphs indexed by discrete time instants; 
            \item \label{item:sketch2} Define \emph{connected equivalent simple} time-evolving graphs by completing a time evolving graph through a set of edges that connect the same nodes at different time instants; 
            \item \label{item:sketch3} Over the connected equivalent simple graph, we can now define, for every time $t$, a graph with Euclidean edges, $G_t$; 
            \item \label{item:sketch4} Define a time-evolving Markov graph to exploit computational advantages.
        \end{enumerate}
    \end{tcolorbox}
    Some comments are in order. Step \ref{item:sketch1} is completely general and does not require any topological structure on every {\em marginal} graph $G_t$, for a given time $t$. Yet, having graphs with Euclidean edges that evolve over time requires some more work, and this fact justifies Step \ref{item:sketch2}, which allows for connectivity, being one of the properties \emph{sine qua non} of a graph with Euclidean edges. Step \ref{item:sketch4} is not mathematically necessary to guarantee the validity of the structure, but it is justified by computational and intuitive reasons as explained throughout. \par
    Step \ref{item:sketch1} starts with a formal definition.
    \begin{definition}[Time-evolving graph]
        \label{def:timeEvGraph}
        Let $T=\curly{0,...,m-1}$ be a (finite) collection of time instants. To every time instant $t \in T$ we associate a simple undirected and weighted graph $G_{t}=(V_{t},E_{t}, w_t)$, with  $V_{t}\cap V_{t'}=\emptyset$ whenever $t\ne t'$. For an edge $e_{t} \in E_{t}$, the corresponding weight is denoted $w(e_{t}):=w_t(e_t)$. We use $n_{t}:= \abs{V_{t}}$ for the number of vertices at time $t$. \par
        Let $G=\curly{G_0,...,G_{m-1}}$ be the associate finite collection of these graphs. Call $V:=\bigcup_t V_t$ the set of vertices, $n:=\abs {V}$ the total number of vertices, and $E_S:=\bigcup_t E_t$ the set of \emph{spatial} edges. Finally, if $v\in V$, whenever convenient we shall write $t(v)$ for the unique value $t$ such that $v\in V_t$. \par
        Let $s:V\to S$ be a mapping from $V$, where $S$ is a set of labels, such that $s(v_1)\ne s(v_2)$ whenever $v_1$ and $v_2$ are two distinct vertices belonging to the same graph $G_t$, $t \in T$. Two vertices $v_1\ne v_2\in V$ are considered the same vertex at different times if $s(v_1)=s(v_2)$. \par
        We call the triple $\boldsymbol{G}=(T, G, s)$ a \emph{time-evolving graph}.
    \end{definition}
    While Definition \ref{def:timeEvGraph} provides a flexible framework to manage graphs that evolve over time, we are going to merge its underlying idea with the one of graph with Euclidean edges presented in Subsection \ref{ssec:GraphsEuclEdges}. Step \ref{item:sketch2} of our routine intends to {\em complete} the time-evolving graph as in Definition  \ref{def:timeEvGraph} so to ensure spatio-temporal connectivity.\par
    Let $\boldsymbol{G}$ be a time-evolving graph. We define its \emph{equivalent simple} time-evolving graph, $\widetilde{\boldsymbol G}=({V}, \widetilde{E})$ as the graph with edges  $\widetilde{E}:=E_S \cup E_T$, with $E_T$ a set of additional edges (called \emph{temporal} edges throughout) that connect the same nodes at different time instants. More precisely, $E_T$ is a subset of $\curly{(v_1,v_2)\in V\times V\,:\,s(v_1)=s(v_2),\,t(v_1)\ne t(v_2)}$. To each new edge $e=(v_1,v_2)\in E_T$ a weight $w(e)>0$ is assigned, while all the other weights remain unchanged. One possibility is to choose $w(e):=\alpha\abs{t(v_1)-t(v_2)}^{-1}$, with $\alpha>0$ a given scale factor. Although we assume this particular expression in all the following examples, we stress that any choice leads to a valid model as long as $w(e) > 0$.\par
\begin{figure}[tp]
        \centering
        \begin{tabular}{cc}
            \includegraphics[width=0.3\textwidth]{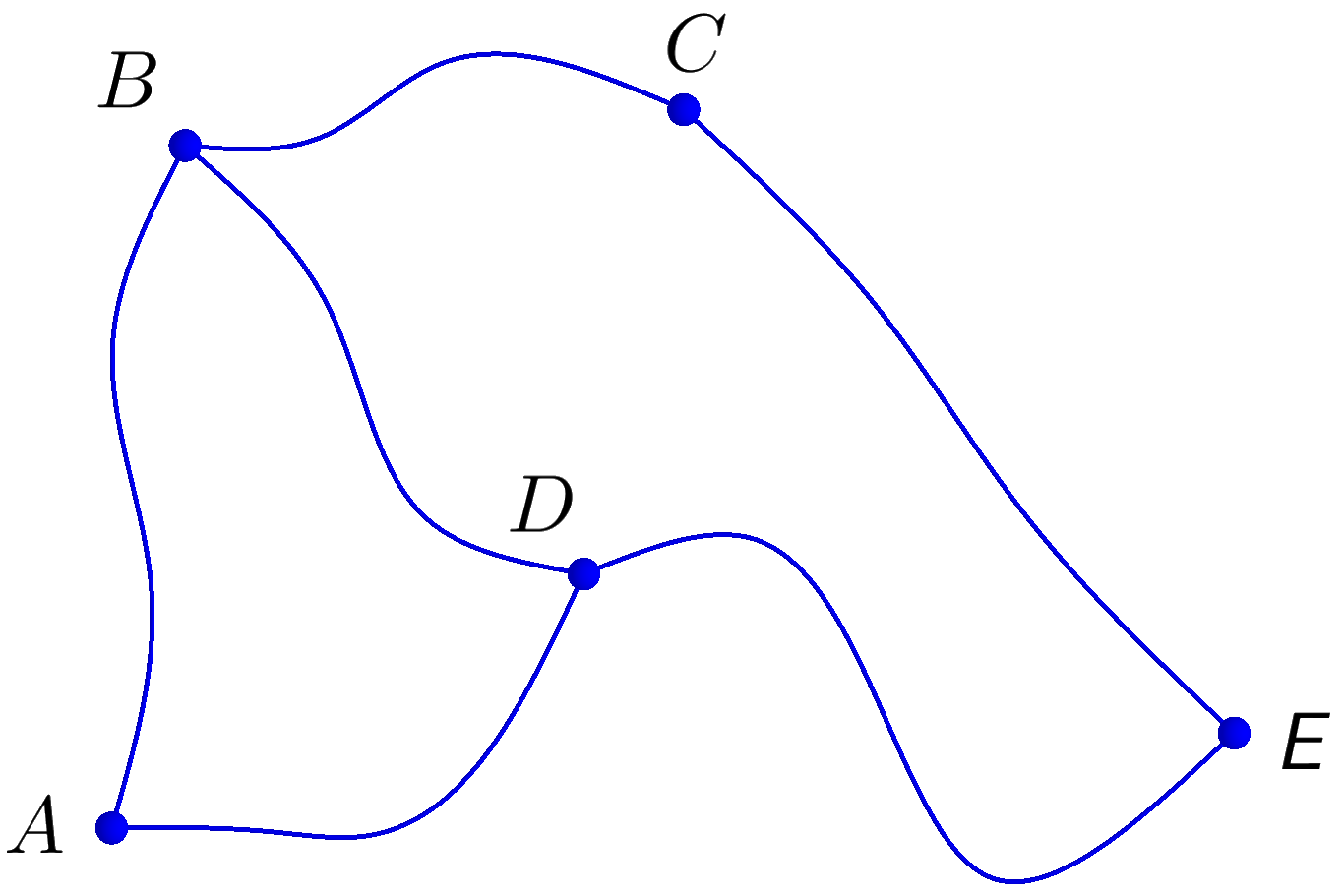} & \includegraphics[width=0.3\textwidth]{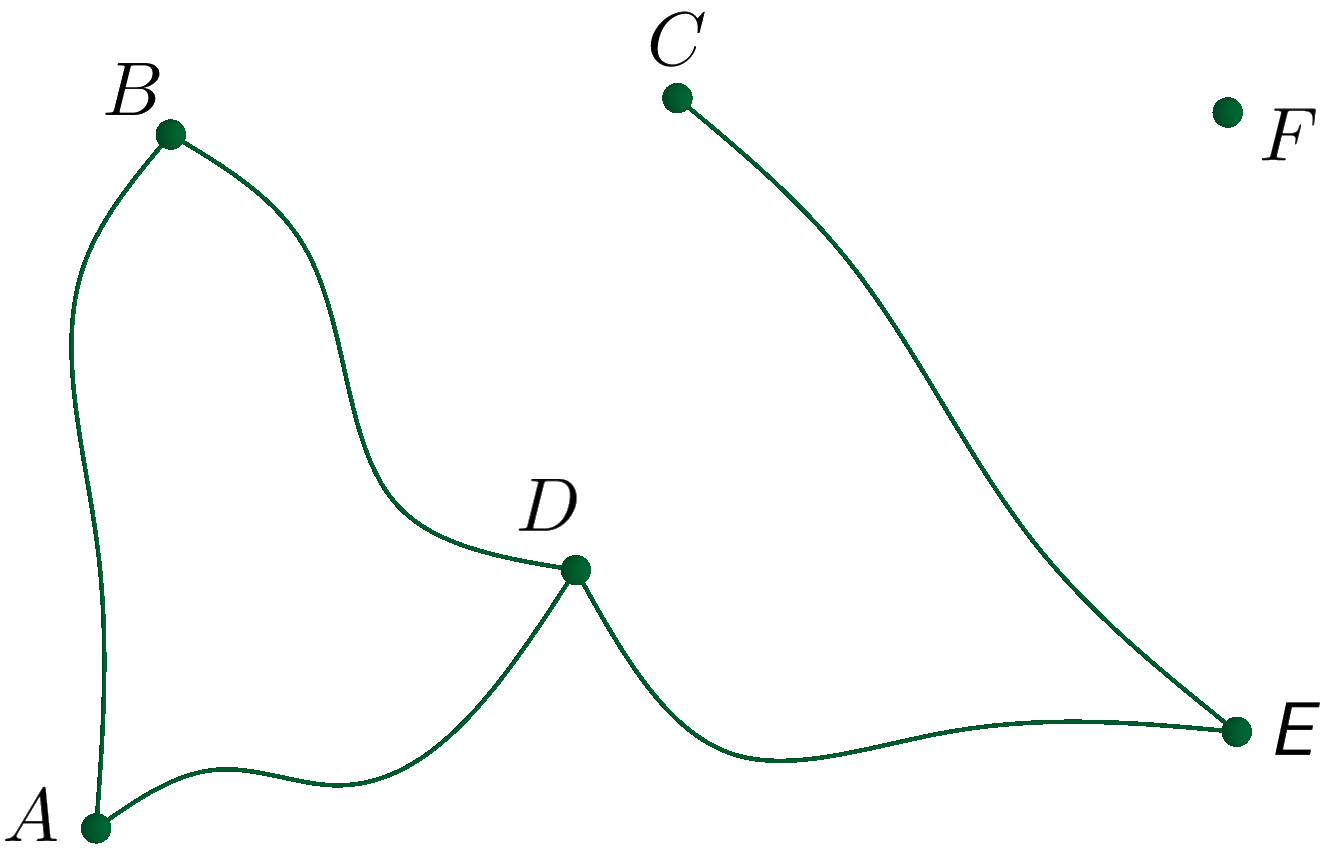}\\
            & \\
            (a) Graph at $t=0$. & (b) Graph at $t=1$. \\
            & \\
            \includegraphics[width=0.3\textwidth]{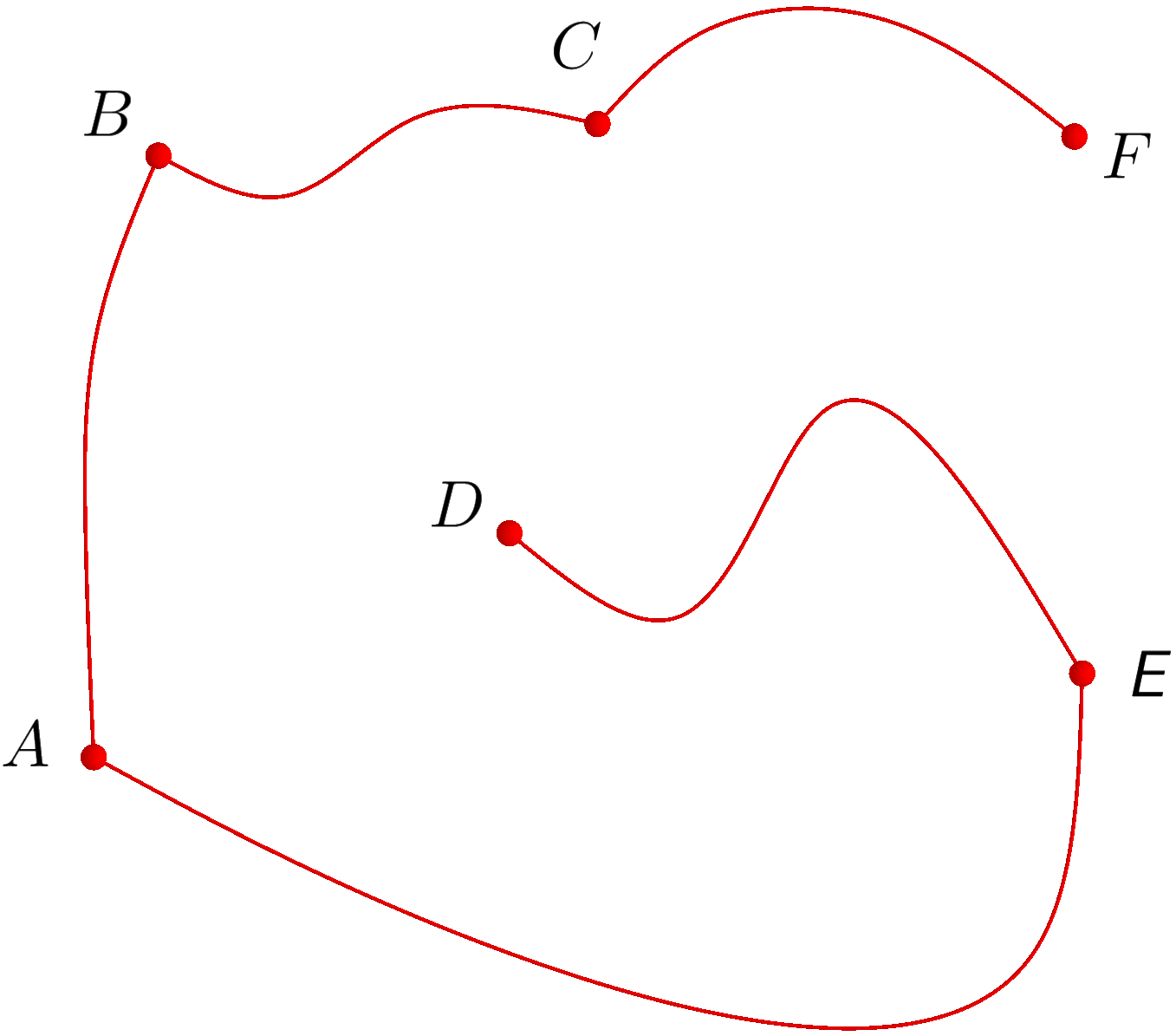} & \includegraphics[width=0.6\textwidth]{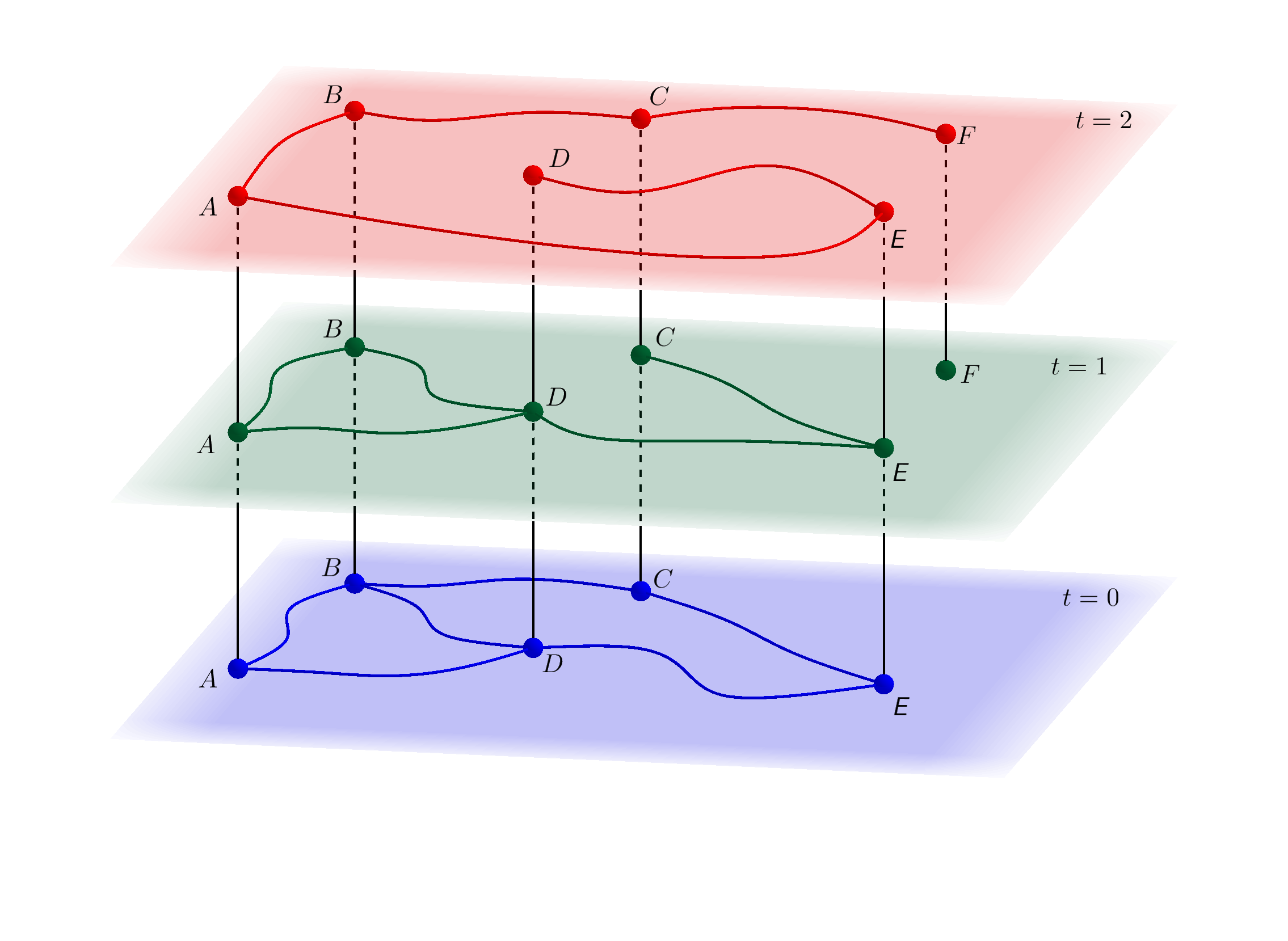} \\
            & \\
            (c) Graph at $t=2$. & (d) Equivalent simple graph. \\
            & \\
        \end{tabular}
        \caption{An example of an equivalent simple graph (bottom-right), with $m=3$, $S=\curly{A,B,C,D,E,F}$, $n_1=5$, $n_2=n_3=6$. The coloured edges belong to $E_S$, whilst the black ones belong to $E_T$. The temporal {\em slices} at time instants $t=0,1,2$ are reported, respectively, at quadrants (a), (b) and (c).}
        \label{fig:LayerGraph}
    \end{figure}

    The intuitive idea behind the construction of an equivalent simple time-evolving graph is to consider $m$ layers, each representing a different temporal instant (namely a graph $G_t$), and connect them by means of additional intra-time edges, which account for the time-dependency of the graphs. Figure \ref{fig:LayerGraph} depicts an example of time-evolving graph and the resulting equivalent simple graph. Henceforth, we will consider each connected component of the equivalent simple graph separately.
    
    Step \ref{item:sketch2} ensures that it becomes feasible to assign, to each temporal label $t \in T$, a graph with Euclidean edges to the equivalent simple graph associated with a given time-evolving graph (Step \ref{item:sketch3}). We note that the choice of the temporal edges needs care. Indeed, the set of possible temporal edges for a fixed label $s \in S$ may grow quadratically in the number of considered temporal instants $m$. \par  
    Our proposal is to connect every node that exists at adjacent times, so that there will be no temporal edge connecting non-adjacent times. Hence, we propose a temporally Markovian structure for the graph (Step \ref{item:sketch4}). This is formalised below.
    \begin{definition}[Time-evolving Markov graph]
        A \emph{time-evolving Markov graph} is a time-evolving graph $\boldsymbol{G}=(V,E)$, where 
        \begin{equation}
            \label{eq:ETMarkovian}
            E_T\subseteq \curly{(v_1,v_2)\in V\times V\,:\, s(v_1)=s(v_2),\, \abs{t(v_1)-t(v_2)}=1}.
        \end{equation}
    \end{definition}
    We stress that temporal Markovianity is not needed to prove the mathematical results following subsequently, which work even for the case of non adjacent layers.\par
    Yet, Markovianity simplifies our job considerably. In fact, it allows for a plain representation of an equivalent simple graph. Furthermore, there are non-negligible computational reasons. Indeed, for large networks, Markovianity allows for sparse Laplacian matrices (block tridiagonal) of the associated equivalent simple graph. This entails huge computational savings in both terms of storage and computation. Finally, allowing edges between non-adjacent layers could lead to a huge number of weights $\alpha$'s. In particular, under the assumption of temporally homogeneous weights (the weight of a temporal edges depends only on the temporal distance between the layers it connects), we would have $m-1$ possible weights if Markovianity is not assumed. For a large collection of time instants, this can become computationally unfeasible. \par
    We call a time-evolving Markov graph $\boldsymbol{G}$ {\em temporally complete} when the set $E_T$ is identically equal to the set in the right hand side of (\ref{eq:ETMarkovian}). Albeit such a property is not required to prove our theoretical results, it is operationally useful as it allows to avoid removing or adding temporal edges. \par

\section{Resistance metrics for linear time}
    \label{sec:resMetricLinearTime}
    We start this section by noting that defining the classical resistance metrics between nodes of the temporally evolving graph is not an issue. Yet, we are dealing with a graph where distances should be computed between any pair of points lying continuously over the edges. This is a major challenge that requires some work as follows. \par
    For the case of static graphs with Euclidean edges, \cite{anderes2020} provide an ingenious construction that allows for a suitable continuously-defined metric on the basis of Brownian bridges and their variograms. \par
    The idea is to follow a similar path, by defining a Gaussian process that is continuously indexed over the edges of an equivalent simple connected graph associated with a given time-evolving graph. \par
    Before going into technical details, we present a brief outline. Following \cite{anderes2020}, we are going to define a distance on all the points of the graph, namely its vertices and the points on its edges. To this aim, we define a Gaussian process $Z$ on every point of the time-evolving equivalent simple Markovian graph and then \emph{define} the distance between two points as the variogram of such process, i.e., for each $u_1,u_2\in \widetilde{\boldsymbol G}$:
    \begin{equation}
        \label{eq:distAsVariogram}
        d(u_1,u_2):=\gamma_Z(u_1,u_2),
    \end{equation} with $\gamma_Z$ as being defined through (\ref{eq:Variogram}). In such a way, we can directly apply Theorem \ref{theo:covarianceConstruction} stated in Appendix \ref{app:isotropicKernelsGeneral} to obtain kernels. Here, $Z:=Z_V+Z_E$ is the sum of two independent Gaussian processes defined on the equivalent simple graph $\widetilde{\boldsymbol G}$. The process $Z_V$ accounts for the structure of the graph (namely its vertices and the weights of its edges) and plays the role of major source of variability (i.e., the distance), whilst $Z_E$ adds some variability on the edges and accounts for the temporal relationship between the same edge at different times. 
    \subsection{Formal construction of \texorpdfstring{$Z_V$}{ZV} and \texorpdfstring{$Z_E$}{ZE}}
        \label{ssec:formalConstrZ_V.Z_E}
        We start by defining the process $Z_V$ through 
        \begin{equation}
            \label{eq:Z_V_linearInterpolation}
            Z_V(u):=(1-\delta_e(u))\,Z_V(\ul u)+\delta_e(u) Z_V(\ol u),
        \end{equation}
        where $u=(\ul u,\ol u, \delta_e(u))$, with $e=(\ul u,\ol u)$ and $\delta_e(u)$ as in Definition \ref{def:graphWithEE}. Further, at the vertices of $\widetilde{\boldsymbol G}$, $Z_V$ is defined as a multivariate normal random variable, denoted $Z_V\big|_{V} \sim \Norm{0}{L^+}$, being $L$ the laplacian matrix associated to the graph $\widetilde{\boldsymbol G}$. The intuitive interpretation is that outside the vertices the process $Z_V$ is obtained through a sheer linear interpolation. \par
        The construction of $Z_E$ is a bit more complex, as $Z_E$ is piecewise defined on a suitable partition of $E$. A formalisation of this concept follows. \par
        For each $e=(v_1,v_2)\in E_S$, we define the \emph{lifespan} of $e$, written $\ls(e)$, as the maximal connected set of time instants, $t$, for which the edge $e$ exists. More formally, $\ls(e)$ is defined as the maximal (with respect to the inclusion partial order) subset of $T$ such that:
        \begin{itemize}
            \item $t(v_1)\in \ls(e)$;
            \item $\ls(e)$ is connected, that is $\forall t_1<t_2\in \ls(e),\,\curly{t_1,\ldots,t_2}\subseteq \ls(e)$;
            \item $\forall t\in \ls(e)$, there exists $ (v_1',v_2')\in E_S$ such that $s(v_1')=s(v_1)$, $s(v_2')=s(v_2)$ and $t(v_1')=t(v_2')=t$.
        \end{itemize}
        
       Figure \ref{fig:LayerGraph} allows to visualise the situation. The lifespan of the edge $(A,B)$ at time $t=0$ is $\curly{0,1,2}$; the lifespan of $(C,E)$ at time $t=1$ is $\curly{0,1}$ and the one of $(B,C)$ at time $t=2$ is $\curly{2}$. \par
        We now define the \emph{life} of $e$ (denoted $\lf(e)$) as the set of edges that represent $e$ at different times and have the same lifespan $\ls(e)$. Formally, we have
        \begin{equation*}
            \lf(e):=\curly{(v_1',v_2')\in E_S: s(v_1')=s(v_1),\,s(v_2')=s(v_2),\,t(v_1')=t(v_2')\in \ls(e)}.
        \end{equation*}
        For convenience, we define the life for temporal edges as well: if $e\in E_T$, $\lf(e):=\curly{e}$.\par 
        It is clear that the set $\curly{\lf(e)\,:\,e\in E_S}$ forms a partition of all the spatial edges $E_S$, and that $\curly{\lf(e)\,:\,e\in E_T}$ is a partition of $E_T$. The main idea is to consider the life of each spatial and temporal edge and define a suitable process on it, being independent from the others. \par
        Let us consider a spatial edge $e\in E_S$ and its lifespan $\ls(e)$. Consider now the set $\ls(e)\times [0,1]$ and define on it a zero-mean Gaussian process $B(t,\delta)$  whose covariance function is given by 
        \begin{equation}
            k_B\round{(t_1,\delta_1),(t_2,\delta_2)}:=k_T(\abs{t_1-t_2})\, k_{BB}(\delta_1,\delta_2), 
        \end{equation}
        with $t_1,t_2\in\ls(e)$ and $\delta_1,\delta_2\in [0,1]$. Here $k_T$ is a temporal kernel defined on $\N$ such that $k_T(0)=1$ and $k_{BB}(\delta_1,\delta_2):=\min(\delta_1,\delta_2)-\delta_1 \delta_2$ is the kernel of the standard Brownian bridge on $[0,1]$. Notice that the spatial marginals of the process $B$ are standard Brownian bridges. We stress that the process $B(t,\delta)$ is only needed for the definition of the process $Z_E$ on $\lf(e)$, as it is better explained below. Now, we define the process $Z_E$ on $\lf(e)$, denoted $Z_E\big |_{\lf(e)}$, as follows: given an edge $e'=(\ul u,\ol u)\in \lf(e)$ and given a point $u=(\ul u,\ol u,\delta)$ on it, 
        \begin{equation} \label{ze}
            Z_E\big |_{\lf(e)}(u):=\sqrt{\ell(e')}\,B(t(\ul u),\delta).
        \end{equation}
        Finally, for each temporal edge $e=[0,\ell(e)]\in E_T$, we define the process $Z_E$ on it as an independent (from both $Z_V$ and $Z_E$ on $E_S$) Brownian bridge on $[0,\ell(e)]$, having covariance function given by $$\Cov{Z_E\big |_e(\delta_1)}{Z_E\big |_e(\delta_2)}=\ell(e)\round{\min(\delta_1,\delta_2)-\delta_1\delta_2}.$$ This concludes the construction of the process on the whole set of the edges.
            
    \subsection{Mathematical properties of the construction}
        \label{ssec:basicProp}
        We remind the reader that the process $Z$ is Gaussian, being the sum of two independent Gaussian processes. Hence, the finite dimensional distribution of $Z$ is completely specified through the second order properties, namely the covariance function. The following result provides an analytical expression for the covariance function associated with $Z$.
        \begin{proposition}
            \label{prop:expressionFor_k_Z}
            Let $u_1,u_2 \in \widetilde{\boldsymbol G}$, with $u_i=(\ul u_i, \ol u_i, \delta_i)$, $i=1,2$. Let $Z=Z_V+Z_E$, with $Z_V$ as defined through (\ref{eq:Z_V_linearInterpolation}) and $Z_E$ as  defined through (\ref{ze}). Then,
            \begin{align}
                \label{eq:covZ}
                k_Z(u_1,u_2)&=\boldsymbol{\delta}_1^\top \, L^+\square{(\ul u_1, \ol u_1),(\ul u_2, \ol u_2)}\,\boldsymbol{\delta}_2\nonumber \\
                &\, + \mathbbm{1}_{\lf(e_1)=\lf(e_2)}\sqrt{\ell(e_1) \ell(e_2)}\,k_T\round{\abs{t(\ul u_1)-t(\ul u_2)}} \round{\min\round{\delta_1, \delta_2}-\delta_1 \delta_2},
            \end{align}
            where $\boldsymbol{\delta}_i:=(1-\delta_i, \delta_i)^\top$, $i=1,2$, and $L^+\square{(\ul u_1, \ol u_1),(\ul u_2, \ol u_2)}$ represents the $2\times 2$ submatrix of $L^+$ with rows $(\ul u_1,\ol u_1)$ and columns $(\ul u_2,\ol u_2)$.
        \end{proposition}
        While noting that this construction is completely general, we also point out that Markovianity properties, whenever aimed, can be achieved through a proper choice of the temporal kernel $k_T$. A reasonable choice for $k_T$ is the correlation function of an autoregressive process of order one, which is given by:
        \begin{equation}
            \label{eq:temporalKernel}
            k_T(h)=\lambda^{\abs{h}}, \qquad h \in \Z,
        \end{equation}
        where $\lambda\in (-1,1)$ is a free parameter and $h\in\Z$ is the lag. \par 
        Notice that the special case $\lambda = 0$, for which $k_T(h)=\mathbbm{1}_{h=0}$, corresponds to the static resistance metric provided by \cite{anderes2020}.

        \begin{figure}[tp]
            \centering
            \begin{subfigure}[b]{0.45\textwidth}
                \begin{adjustbox}{width=\linewidth}
                    \begin{tikzpicture}[font=\footnotesize]
                        \begin{axis}
                        [
                            samples = 200,
                            domain = 0:1,
                            samples y=0, 
                            ytick = {0,1,2,3},
                            xtick = {0, 0.5, 1},
                            zmin = -1, zmax = 1,
                            xlabel=$\delta$, ylabel=$t$,  
                            view={15}{30},
                            plot box ratio={1}{2.5}{1},
                            title=\text{$\lambda=0.75$}
                        ]
                        \addplot3 [fill opacity=0.25,
                                draw=blue!60!black,thick,
                                fill=blue!60!black,
                                mark=none,] table [x ="x", y expr=0, z="time1"] {\brownianBridgeDataHighCorr};
                        \addplot3 [fill opacity=0.25,
                                draw=green!60!black,thick,
                                fill=green!60!black,
                                mark=none,] table [x ="x", y expr=1, z="time2"] {\brownianBridgeDataHighCorr};
                        \addplot3 [fill opacity=0.25,
                                draw=red!80!black,thick,
                                fill=red!80!black,
                                mark=none,] table [x ="x", y expr=2, z="time3"] {\brownianBridgeDataHighCorr};
                        \addplot3 [fill opacity=0.25,
                                draw=orange!80!black,thick,
                                fill=orange!80!black,
                                mark=none,] table [x ="x", y expr=3, z="time4"] {\brownianBridgeDataHighCorr};
                        \end{axis}
                    \end{tikzpicture}
                \end{adjustbox}
            \end{subfigure}
            \hfill
            \begin{subfigure}[b]{0.45\textwidth}
                \begin{adjustbox}{width=\linewidth}
                    \begin{tikzpicture}[font=\footnotesize]
                        \begin{axis}
                        [
                            samples = 200,
                            domain = 0:1,
                            samples y=0, 
                            ytick = {0,1,...,3},
                            xtick = {0, 0.5, 1},
                            zmin = -1, zmax = 1,
                            xlabel=$\delta$, ylabel=$t$,  
                            view={15}{30},
                            plot box ratio={1}{2.5}{1},
                            title=\text{$\lambda=0.25$}
                        ]
                        \addplot3 [fill opacity=0.25,
                                draw=blue!60!black,thick,
                                fill=blue!60!black,
                                mark=none,] table [x ="x", y expr=0, z="time1"] {\brownianBridgeDataLowCorr};
                        \addplot3 [fill opacity=0.25,
                                draw=green!60!black,thick,
                                fill=green!60!black,
                                mark=none,] table [x ="x", y expr=1, z="time2"] {\brownianBridgeDataLowCorr};
                        \addplot3 [fill opacity=0.25,
                                draw=red!80!black,thick,
                                fill=red!80!black,
                                mark=none,] table [x ="x", y expr=2, z="time3"] {\brownianBridgeDataLowCorr};
                        \addplot3 [fill opacity=0.25,
                                draw=orange!80!black,thick,
                                fill=orange!80!black,
                                mark=none,] table [x ="x", y expr=3, z="time4"] {\brownianBridgeDataLowCorr};
                        \end{axis}
                    \end{tikzpicture}
                \end{adjustbox}
            \end{subfigure}
            
            \begin{subfigure}[b]{0.45\textwidth}
                \begin{adjustbox}{width=\linewidth}
                    \begin{tikzpicture}[font=\footnotesize]
                        \begin{axis}
                        [
                            samples = 200,
                            domain = 0:1,
                            samples y=0, 
                            ytick = {0,...,3},
                            xtick = {0, 0.5, 1},
                            zmin = -1, zmax = 1,
                            xlabel=$\delta$, ylabel=$t$,  
                            view={15}{30},
                            plot box ratio={1}{2.5}{1},
                            title=\text{$\lambda=0.00$}
                        ]
                        \addplot3 [fill opacity=0.25,
                                draw=blue!60!black,thick,
                                fill=blue!60!black,
                                mark=none,] table [x ="x", y expr=0, z="time1"] {\brownianBridgeDataZeroCorr};
                        \addplot3 [fill opacity=0.25,
                                draw=green!60!black,thick,
                                fill=green!60!black,
                                mark=none,] table [x ="x", y expr=1, z="time2"] {\brownianBridgeDataZeroCorr};
                        \addplot3 [fill opacity=0.25,
                                draw=red!80!black,thick,
                                fill=red!80!black,
                                mark=none,] table [x ="x", y expr=2, z="time3"] {\brownianBridgeDataZeroCorr};
                        \addplot3 [fill opacity=0.25,
                                draw=orange!80!black,thick,
                                fill=orange!80!black,
                                mark=none,] table [x ="x", y expr=3, z="time4"] {\brownianBridgeDataZeroCorr};
                        \end{axis}
                    \end{tikzpicture}
                \end{adjustbox}
            \end{subfigure}
            \hfill
            \begin{subfigure}[b]{0.45\textwidth}
                \begin{adjustbox}{width=\linewidth}
                    \begin{tikzpicture}[font=\footnotesize]
                        \begin{axis}
                        [
                            samples = 200,
                            domain = 0:1,
                            samples y=0, 
                            ytick = {0,1,...,3},
                            xtick = {0, 0.5, 1},
                            zmin = -1, zmax = 1,
                            xlabel=$\delta$, ylabel=$t$,  
                            view={15}{30},
                            plot box ratio={1}{2.5}{1},
                            title=\text{$\lambda=-0.75$}
                        ]
                        \addplot3 [fill opacity=0.25,
                                draw=blue!60!black,thick,
                                fill=blue!60!black,
                                mark=none,] table [x ="x", y expr=0, z="time1"] {\brownianBridgeDataNegativeCorr};
                        \addplot3 [fill opacity=0.25,
                                draw=green!60!black,thick,
                                fill=green!60!black,
                                mark=none,] table [x ="x", y expr=1, z="time2"] {\brownianBridgeDataNegativeCorr};
                        \addplot3 [fill opacity=0.25,
                                draw=red!80!black,thick,
                                fill=red!80!black,
                                mark=none,] table [x ="x", y expr=2, z="time3"] {\brownianBridgeDataNegativeCorr};
                        \addplot3 [fill opacity=0.25,
                                draw=orange!80!black,thick,
                                fill=orange!80!black,
                                mark=none,] table [x ="x", y expr=3, z="time4"] {\brownianBridgeDataNegativeCorr};
                        \end{axis}
                    \end{tikzpicture}
                \end{adjustbox}
            \end{subfigure}
            
            \caption{Draws from the process $Z_E$ on an edge with lifespan $\curly{0,1,2,3}$, for several values of the parameter $\lambda$.}
            \label{fig:corrBrownianBridges}
        \end{figure}
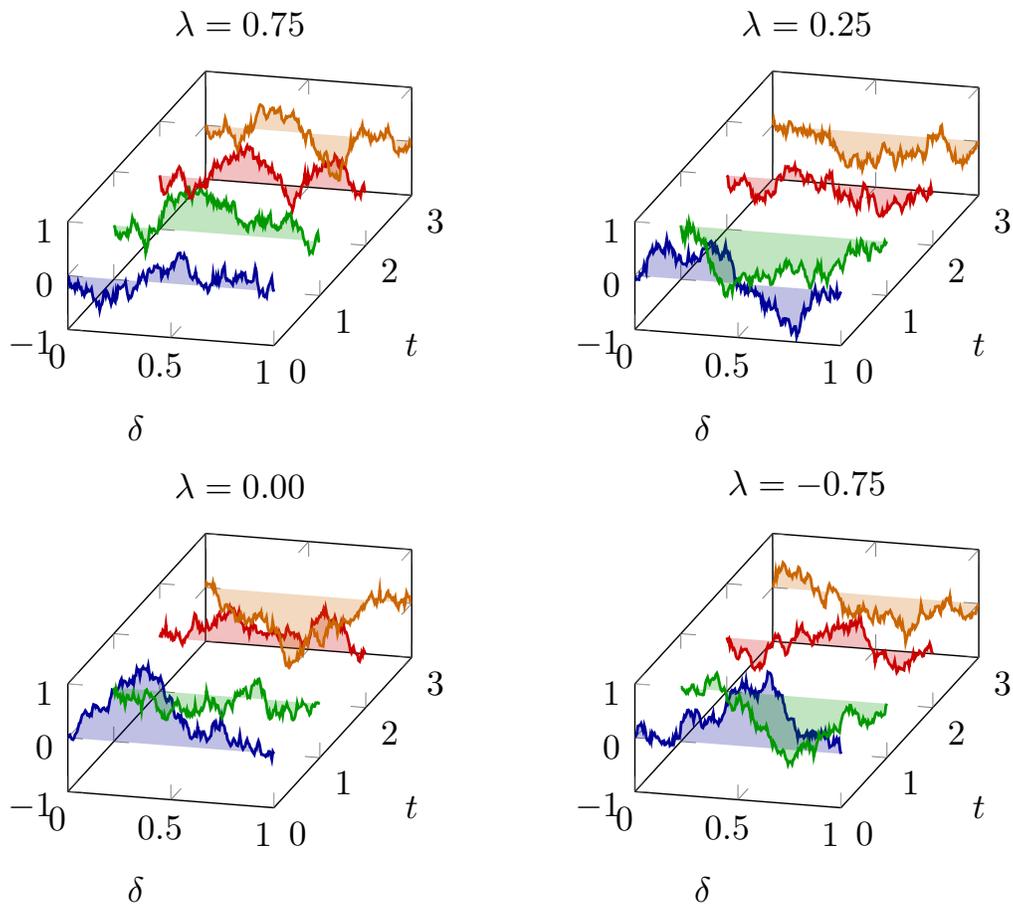

        Figure \ref{fig:corrBrownianBridges} depicts some realisations for the process $Z_E$ over an edge for different values of the parameter $\lambda$. The parameter $\lambda$ for the edges plays a similar role of the parameter $\alpha$ for the nodes: their are both closely related to the inter-dependency of the process at different times. Indeed, $\lambda$ measures how much the process $Z_E$ is correlated between two times $t_1,t_2\in \ls(e)$. Analogously, the value $\alpha$ as weight of the temporal edge $e\in E_T$, is related to the partial correlation of the endpoints of $e$ given everything else. As a consequence, it is natural to choose a high value of $\lambda$ for high values of $\alpha$ and vice-versa. We notice that it is natural to choose non-negative values for $\lambda$, as we usually expect a non-negative correlation between the values of $Z_E$ for close times.\par
        Using equation (\ref{eq:distAsVariogram}), we get the following expression for the distance between any to points $u_1,u_2 \in \widetilde{\boldsymbol G}$.
        \begin{equation}
            \label{eq:distFromKernel}
            d(u_1,u_2)=k_Z(u_1,u_1)+k_Z(u_2,u_2)-2 k_Z(u_1,u_2).
        \end{equation}
        The formal statement below provides a complete description of the space $(\widetilde{\boldsymbol G},d)$, that is, the time-evolving graph $\widetilde{\boldsymbol G}$ equipped with the metric $d$.
        \begin{proposition}
            \label{prop:semiMetric}
            Let $d$ be the mapping defined at (\ref{eq:distFromKernel}). Then, the pair
            $(\widetilde{\boldsymbol G}, d)$ is a semi-metric space. 
        \end{proposition}
        One might ask whether a stronger assertion holds for the pair $(\widetilde{\boldsymbol G}, d)$ as defined above. The next statement provides a negative answer.

        \begin{figure}[tp]
            \centering
            \resizebox{0.5\textwidth}{!}{
                \begin{tikzpicture}
                    \node (A) at (-1,0) {$A_0$};
                    \node (B) at (3,0) {$B_0$};
                    \node (C) at (0,2) {$A_1$};
                    \node (D) at (2,2) {$B_1$};
                    \node (E) at (0.5,4) {$A_2$};
                    \node (F) at (1.5,4) {$B_2$};
                    \node (t0) at (4,0) {$t=0$};
                    \node (t1) at (4,2) {$t=1$};
                    \node (t2) at (4,4) {$t=2$};
    
                    \filldraw (1,0) circle[radius=1.5pt];
                    \filldraw (1,2) circle[radius=1.5pt];
                    \filldraw (1,4) circle[radius=1.5pt];
                    
                    \node[above=3pt, outer sep=0pt] at (1,0) {$P$};
                    \node[above=3pt, outer sep=0pt] at (1,2) {$Q$};
                    \node[above=3pt, outer sep=0pt] at (1,4) {$R$};
                
                    \path [-] (A) edge (B);
                    \path [-] (A) edge (C);
                    \path [-] (B) edge (D);
                    \path [-] (C) edge (D);
                    \path [-] (C) edge (E);
                    \path [-] (D) edge (F);
                    \path [-] (E) edge (F);
                \end{tikzpicture}
            }
            \caption{An example of equivalent simple graph for which the semi-distance defined at (\ref{eq:distFromKernel}) does not satisfy the triangle inequality.}
            \label{fig:triangleIneqCounterExample}
        \end{figure}
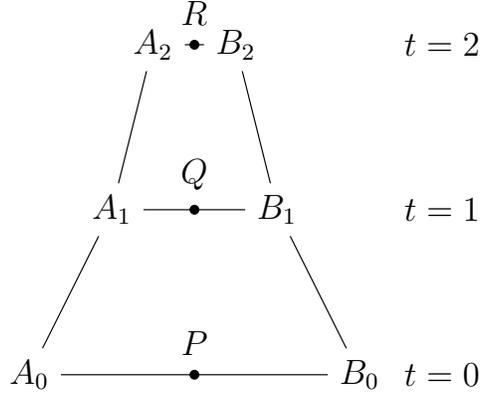

A counterexample is given by the graph depicted in Figure \ref{fig:triangleIneqCounterExample}, where the length of the top edge $(A_2,B_2)$ becomes vanishingly small, while length of the bottom edge $(A_0,B_0)$ grows to infinity. For such a graph, we have $d(P,Q)+d(Q,R) < d(P,R)$ (see the proof of Proposition \ref{prop:notAMetric} in the Appendix for more details).
        \begin{proposition}
            \label{prop:notAMetric}
           Let $d$ be the mapping defined at (\ref{eq:distFromKernel}). Then, the pair $(\widetilde{\boldsymbol G}, d)$ is not a metric space.
        \end{proposition}
        
        \begin{remark}
            Although in general our extension to the classic resistance distance is not a metric, it retains some of its properties.
            \begin{enumerate}
                \item $(V, d\big|_V)$ is a metric space.
                \item For all $t$, $(G_t,d\big|_{G_t})$ coincides with the restriction on $G_t$ of the resistance metric of \cite{anderes2020} computed on the whole graph $\widetilde{\boldsymbol G}$, hence it is a metric and it is invariant to splitting edges and merging edges at degree 2 vertices \citep[Propositions 2 and 3]{anderes2020}. 
            \end{enumerate}
        \end{remark}

    \subsection{A special case: time-evolving linear networks}
        \cite{anderes2020} defined graphs with Euclidean edges as a generalisation of linear networks, and Euclidean trees with a given number of leaves. For both cases, edges are linear. This case is not especially exciting for the framework proposed in this paper. The reason is that a simple isometric embedding arguments as in \cite{tang} proves that one can embed a time-evolving linear network in $\R \times \R^2=\R^3$, where the first component indicates time. As a consequence, it is immediate to build a vast class of covariance functions on a time-evolving linear network by a sheer restriction of a given covariance function defined on $\R^3$. However, such a method does not take into account the \emph{structure} of the graph: two points that are close in $\R^3$ but far in the time-evolving graph could have a high correlation. Section \ref{sec:reproducingKernels} illustrates how to build kernels over the special topologies proposed in this paper. Apparently, the choices are more restrictive than the ones available for the case of linear networks, but they ensure that the spatio-temporal structure is taken into account.

\section{Circular time and periodic graphs}
    \label{sec:resMetricPeriodicTime}
    Perhaps the main drawback of using the resistance distance in the layer graphs that express the spatio-temporal variability is that, when adding one or more new layers, the distances between the points of the previous layers may change (more specifically: they may decrease). Indeed, whenever new paths between a couple of points are added, the effective distance between such points decreases, as the current meets less resistance.  This presents a critical interpretation problem: for a given time series,  let  new data be added on a daily basis. Then, inference routines may provide different results when compared to the results of same inference techniques applied to the updated time series. Indeed, as the distances may vary, the covariances between the same space-time points may vary as well.\par    
    Here, we consider the alternative of time-evolving \emph{periodic} networks, \emph{i.e.} time-evolving networks whose evolution repeats after a fixed amount of time instants (number of layers). Not only does this construction solve the above-mentioned issue, but it also suits many phenomena whose evolution present both linear and periodic components.\par
    
    \begin{definition}[Time-evolving periodic graph]
        Let $\boldsymbol{G}=\curly{G_0, G_1, \dots}$ be a countable sequence of graphs. Then, $\boldsymbol{G}$ is a \emph{time-evolving periodic graph} if there exists a natural number $m\geq 3$ such that, for all $t\in \N$, $G_t=G_{t+m}$. Its equivalent simple Markovian periodic graph $\widetilde{\boldsymbol G}$ is built by connecting $G_0,\dots,G_{m-1}$ through the set of edges
        \begin{equation}
            \label{eq:E_TforPeriodicGraphs}
            E_T=\curly{(v_1,v_2)\in V\times V\,:\,s(v_1)=s(v_2),\,\abs{t(v_1)-t(v_2)}\equiv \pm 1\pmod{m}}.
        \end{equation}
        Each point in the resulting space-time is denoted by its true time $t\in\R_0^+$, by the endpoints of the edge $e$ it lies on and by the relative distance $\delta_e(u)$ from the first one: we write $u=(t, \ul u, \ol u, \delta_e(u))$, where $e=(\ul u, \ol u)$. Notice that $t\in\N$ whenever $u$ belongs to a temporal layer, while $t$ is not integer if $u$ belongs to the inner part of a temporal edge $e\in E_T$. Given a point $u\in V \cup \bigcup E_S$, we sometimes write $\tau(u)$ as the unique layer $\tau\in\curly{0,\ldots,m-1}$ that contains $u$. Clearly $\tau(u)\equiv t(u) \pmod{m}$.
    \end{definition}
    We start by noting that the previously-mentioned issue about linear time-evolving graphs is overcome by this construction. 
    Indeed, once the full periodic structure has been established, the Laplacian matrix needs be computed only once, regardless of how many new time points are added. \par
    A second remark comes from the metric construction, which necessarily needs to be adapted to a periodic process. Otherwise, some counter-intuitive properties can arise. Suppose the distance $d(u_1,u_2)$ is defined as in (\ref{eq:distAsVariogram}). Then, for  any couple of points $u_1=(t_1, \ul u_1, \ol u_1, \delta_e(u_1))$ and $u_2=(t_2, \ul u_2, \ol u_2, \delta_e(u_2))$ with $\ul u_1 = \ul u_2$, $\ol u_1 = \ol u_2$ and $\delta_e(u_1)=\delta_e(u_2)$, even when $t_1 \ne t_2$, the distance would be identically equal to zero. Hence, a different definition for the process $Z$ is necessary. \par
    For a point $u=(t, \ul u, \ol u, \delta_e(u))\in \widetilde{\boldsymbol G}$, we define the process $Z$ for the periodic graph as follows:
    \begin{equation}
        \label{eq:ZforPeriodicCase}
        Z(u):=Z_V(u)+Z_E(u)+\beta W(t),
    \end{equation}
    where $\beta>0$ is a given parameter, $W$ is a standard Wiener process, $Z_V$ is the same process as in the linear-time case, while $Z_E$, albeit similar, presents some difference with respect to the construction given in Subsection \ref{ssec:formalConstrZ_V.Z_E}, aimed to capture the time structure of the periodic graph. \par
    Let $e=(v_1,v_2)\in E_S$: we define the \emph{lifespan} of $e$ as the maximal subset of $\curly{0,\ldots,m-1}$ such that:
    \begin{itemize}
        \item $\tau(v_1)\in\ls(e)$,
        \item $\ls(e)$ is connected, i.e. if $\tau_1<\tau_2\in\ls(e)$, then $\curly{\tau_1,\ldots,\tau_2}\subseteq\ls(e)$ or $\curly{\tau_2,\ldots, m-1, 0, \ldots, \tau_1}\subseteq\ls(e)$,
        \item $\forall \tau\in \ls(e),\,\exists (v_1',v_2')\in E_S$ such that $s(v_1')=s(v_1)$, $s(v_2')=s(v_2)$ and $\tau(v_1')=\tau(v_2')=\tau$.
    \end{itemize}

    \begin{figure}[tp]
        \centering
        \includegraphics[width=0.7\textwidth]{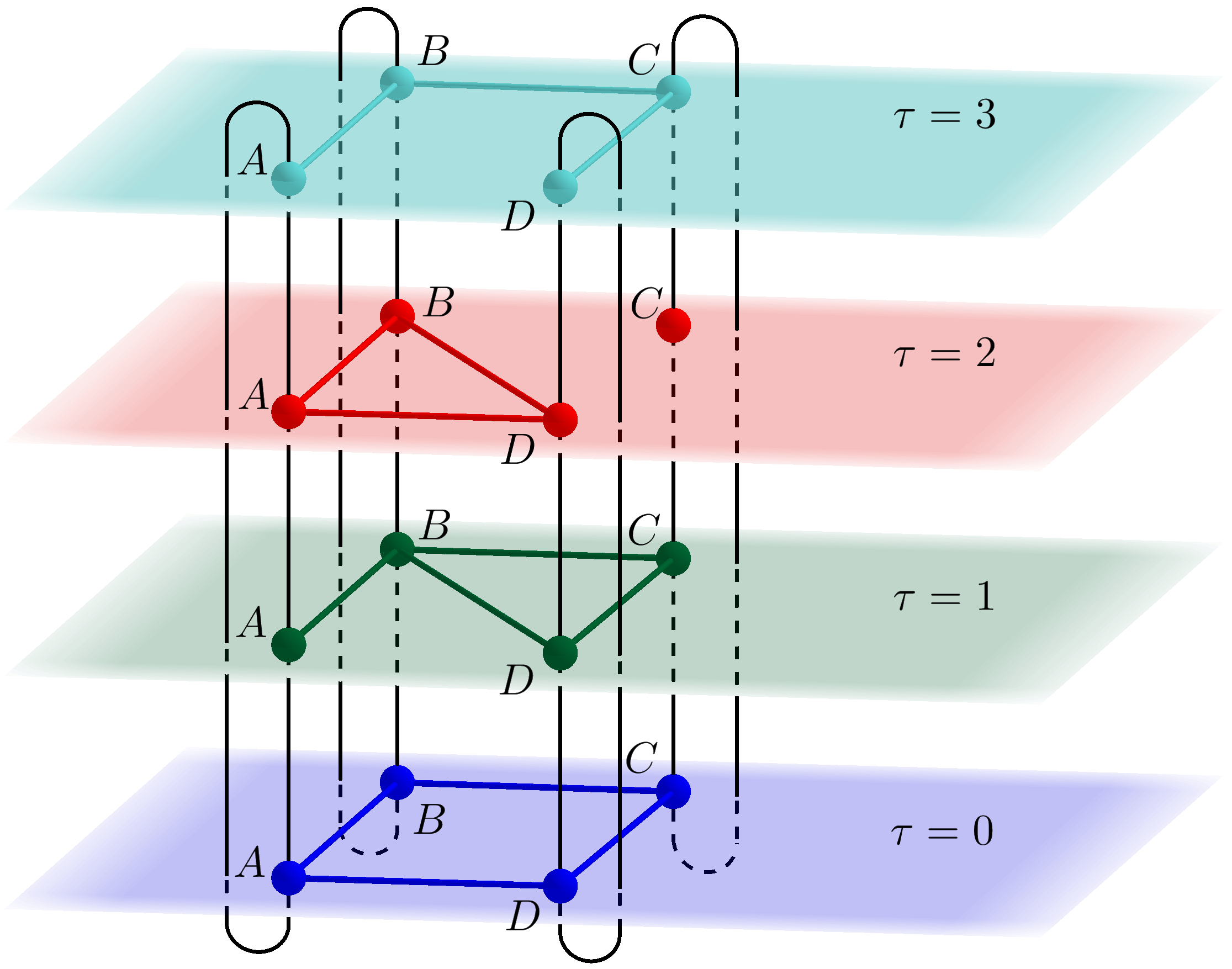}
        \caption{An example of an equivalent simple graph for a periodic time-evolving graph with $m=4$ and $S =\curly{A,B,C,D}$. The coloured edges belong to $E_S$, whilst the black ones belong to $E_T$.}
        \label{fig:LayerGraphPeriodic}
    \end{figure}    
    Figure \ref{fig:LayerGraphPeriodic} depicts this situation. Here, the lifespan of the edge $(A,B)$ at time $\tau=0$ is $\curly{0,1,2,3}$; the lifespan of $(A,D)$ at time $\tau=2$ is $\curly{2}$; the lifespan of $(C,D)$ at time $\tau=3$ is $\curly{0,1,3}$.\par 
    The definition of life of any edge $e\in E$ remains unchanged: if $e\in E_S$,
    \begin{equation*}
        \lf(e):=\curly{(v_1',v_2')\in E_S: s(v_1')=s(v_1),\,s(v_2')=s(v_2),\,\tau(v_1')=\tau(v_2')\in \ls(e)}
    \end{equation*}
    while, if $e\in E_T$, $\lf(e):=\curly{e}$.\par
    The definition of $Z_E$ is now identical to the one of the linear-time graph, exception made for the choice of the temporal kernel $k_T$. Indeed, we ought to consider that the time is now cyclic in the dependence structure of the temporal layers $\tau\in\curly{0,...,m-1}$. It is reasonable to model the process underlying the temporal kernel $k_T$ by means of a graphical model, as it embodies the idea of conditional independence. For a given spatial edge $e\in E_S$, we distinguish two cases: whether the lifespan of $e$ is the whole temporal set $T=\curly{0,...,m-1}$ or not.

    \subsection{The lifespan coincides with \texorpdfstring{$T$}{T}}
        \label{ssec:circTime_lifespanFull}
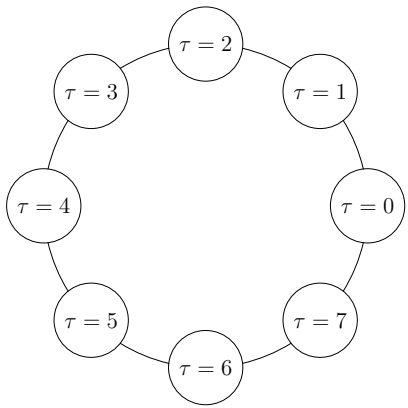
\begin{figure}[tp]
            \centering
            \resizebox{5.5 cm}{!}{
                \begin{tikzpicture}
                    \def \m {8}
                    \def \radius {3cm}
                    \def \margin {13}  
                    
                    \foreach \s in {1,...,\m} {
                        \edef \y {\number \numexpr \s - 1 \relax}
                        \node[draw, circle] at ({360/\m * (\s - 1)}:\radius) {$\tau=\y$};
                        \draw[-, >=latex] ({360/\m * (\s - 1)+\margin}:\radius) 
                            arc ({360/\m * (\s - 1)+\margin}:{360/\m * (\s)-\margin}:\radius);
                    }
                \end{tikzpicture}
            }
            \caption{Conditional dependence structure of the process $Z_T$ for $m=8$.}
            \label{fig:circulantGraph8}
        \end{figure}

        In this case, we define the covariance matrix of a zero-mean Gaussian random vector $Z_T:\curly{0,...,m-1}\to\R$ via its precision matrix. More precisely, let $G_T$ be the circulant graph with $m$ nodes (labelled by $\tau\in\curly{0,...,m-1}$) and $m$ edges between adjacent nodes, as shown in Figure \ref{fig:circulantGraph8}. 
        We associate each edge with a given weight $\rho\in \big[0, \frac 1 2 \big)$, which represents the partial correlation between subsequent times. As a consequence, the precision matrix $\Theta_{Z_T}$ is the circulant matrix that follows. 
        \begin{equation*}
            \Theta_{Z_T}=\kappa \begin{bmatrix}
                1       &   -\rho   &   0       &   \ldots  &   0       &   -\rho   \\
                -\rho   &   1       &   -\rho   &   \ldots  &   0       &   0       \\
                0       &   -\rho   &   1       &   \ldots  &   0       &   0       \\
                \vdots  &   \vdots  &   \vdots  &   \ddots  &   \vdots  &   \vdots  \\
                0       &   0       &   0       &   \ldots  &   1       &   -\rho   \\
                -\rho   &   0       &   0       &   \ldots  &   -\rho   &   1
            \end{bmatrix}
        \end{equation*}
        Here, $\kappa>0$ is a normalising constant which role is to make the covariance matrix $\Sigma_{Z_T}:=\round{\Theta_{Z_T}}^{-1}$ a correlation matrix (namely the variances of every entry of $Z_T$ should be 1). Notice that the matrix $\Sigma_{Z_T}$ is a symmetric circulant matrix: as a consequence, it is possible to store only its first column, which will be denoted by $\sigma_{Z_T}\in \R^m$. In Figure \ref{fig:example_k_T}, the values of the vector $\sigma_{Z_T}$ are plotted for some values of $m$ and $\rho$. 
        \begin{figure}[tp]
            \centering
            \begin{subfigure}[b]{0.45\textwidth}
                \centering
                \begin{tikzpicture}
                    \begin{scriptsize}
                    \begin{axis}[
                        xtick={0,1,2,3,4,5,6,7,8},
                        xticklabels={0,1,2,3,4,5,6,7,0},
                        ymin = -0.2, ymax = 1.5,
                        width = \textwidth,
                        height = \textwidth,
                        xlabel = $\abs{\tau_1-\tau_2}$,
                        ylabel = $\Corr{Z_T(\tau_1)}{Z_T(\tau_2)}$
                        ]
                        \addplot[color=green,]
                        coordinates {(0,1) (1,0.6493) (2,0.4429) (3,0.335) (4,0.3015) (5,0.335) (6,0.4429) (7,0.6493) (8,1)};
                        \addlegendentry{$\rho=0.45$}
                        \addplot[color=blue,]
                        coordinates {(0,1) (1,0.5058) (2,0.2646) (3,0.1556) (4,0.1245) (5,0.1556) (6,0.2646) (7,0.5058) (8,1)};
                        \addlegendentry{$\rho=0.4$}
                        \addplot[color=red,]
                        coordinates {(0,1) (1,0.2087) (2,0.0436) (3,0.0095) (4,0.0038) (5,0.0095) (6,0.0436) (7,0.2087) (8,1)};
                        \addlegendentry{$\rho=0.2$}
                    \end{axis}
                    \end{scriptsize}
                \end{tikzpicture}
                \caption*{The correlations of $Z_T$ for some values of $\rho$ and $m=8$.}
                \label{fig:corrZ_T_m8}
            \end{subfigure}
            \hfill
            \begin{subfigure}[b]{0.45\textwidth}
                \centering
                \begin{tikzpicture}
                    \begin{scriptsize}
                    \begin{axis}[
                        xtick={0,5,10,15,20},
                        xticklabels={0,5,10,15,0},
                        ymin = -0.2, ymax = 1.5,
                        width = \textwidth,
                        height = \textwidth,
                        xlabel = $\abs{\tau_1-\tau_2}$,
                        ylabel = $\Corr{Z_T(\tau_1)}{Z_T(\tau_2)}$
                        ]
                        \addplot[color=green,]
                        coordinates {(0,1) (1,0.6269) (2,0.3931) (3,0.2466) (4,0.1549) (5,0.0976) (6,0.0621) (7,0.0403) (8,0.0275) (9,0.0208) (10,0.0187) (11,0.0208) (12,0.0275) (13,0.0403) (14,0.0621) (15,0.0976) (16,0.1549) (17,0.2466) (18,0.3931) (19,0.6269) (20,1)};
                        \addlegendentry{$\rho=0.45$}
                        \addplot[color=blue,]
                        coordinates {(0,1) (1,0.5) (2,0.25) (3,0.125) (4,0.0625) (5,0.0313) (6,0.0157) (7,0.0079) (8,0.0042) (9,0.0024) (10,0.002) (11,0.0024) (12,0.0042) (13,0.0079) (14,0.0157) (15,0.0313) (16,0.0625) (17,0.125) (18,0.25) (19,0.5) (20,1)};
                        \addlegendentry{$\rho=0.4$}
                        \addplot[color=red,]
                        coordinates {(0,1) (1,0.2087) (2,0.0436) (3,0.0091) (4,0.0019) (5,4e-04) (6,1e-04) (7,0) (8,0) (9,0) (10,0) (11,0) (12,0) (13,0) (14,1e-04) (15,4e-04) (16,0.0019) (17,0.0091) (18,0.0436) (19,0.2087) (20,1)};
                        \addlegendentry{$\rho=0.2$}
                    \end{axis}
                    \end{scriptsize}
                \end{tikzpicture}
                \caption*{The correlations of $Z_T$ for some values of $\rho$ and $m=20$.}
                \label{fig:corrZ_T_m20}
            \end{subfigure}
            
            \caption{Some examples of the correlation functions $k_T$ in the case $\ls(e)=\curly{0,\dots,m-1}$.}
            \label{fig:example_k_T}
        \end{figure}
                
    \subsection{The lifespan does not coincide with \texorpdfstring{$T$}{T}}    
        \label{ssec:circTime_lifespanInterrupted}
        In this case, the life of the edge $e$ is interrupted. Thus, it is reasonable to consider the different parts of the life of $e$ as independent. To this aim, we consider the subgraph of $G_T$ that represents the evolution of the edge $e$. More precisely, we remove from $G_T$ all the nodes $\tau$ for which the edge $e$ does not exist and we remove from $G_T$ all the edges whose at least one endpoint has been eliminated. Next, we consider all the connected components of the so-obtained graph and define an autoregressive model on each of them, independently from the others (similarly to the linear case of Section \ref{ssec:basicProp}). More precisely, we define the covariance matrix of the process $Z_T$ as a block-diagonal matrix whose diagonal blocks are of the form
        \begin{equation*}
            \begin{bmatrix}
                1   &   \lambda    &   \ldots  &   \lambda^{j-1}\\
                \lambda  &   1   &   \ldots  &   \lambda^{j-2}\\
                \vdots  &   \vdots  &   \ddots  &   \vdots\\
                \lambda^{j-1}    &   \lambda^{j-2}    &   \ldots  & 1
            \end{bmatrix},
        \end{equation*}
        being $\lambda\in (-1,1)$ the lag-1 correlation and $j$ the number of times $\tau$ that belong to $\ls(e)$, \textit{i.e.} $j:=\abs{\ls(e)}$.

    \subsection{Second-order properties of \texorpdfstring{$Z$}{Z} in the circular case}
        The following result illustrates the analytic expression for the covariance function associated with $Z$ in the construction of the metric associated with $\widetilde{\boldsymbol G}$ in the periodic case. 
        \begin{proposition}
            \label{prop:expressionFor_k_Z_Periodic}
            Let $u_1=(t_1, \ul u_1, \ol u_1, \delta_e(u_1))\in \widetilde{\boldsymbol G}$ and $u_2=(t_2, \ul u_2, \ol u_2, \delta_e(u_2))\in \widetilde{\boldsymbol G}$. Then the kernel of the process $Z$ defined on $\widetilde{\boldsymbol G}$ enjoys the following representation:
            \begin{align}
                \label{eq:covZ_Periodic}
                k_Z(u_1,u_2)&=\boldsymbol{\delta}_1^\top \, L^+\square{(\ul u_1, \ol u_1),(\ul u_2, \ol u_2)}\,\boldsymbol{\delta}_2\nonumber \\
                &\quad + \mathbbm{1}_{\lf(e_1)=\lf(e_2)}\sqrt{\ell(e_1) \ell(e_2)}\,k_T\round{\tau(\ul u_1),\tau(\ul u_2)} \round{\min\round{\delta_1, \delta_2}-\delta_1 \delta_2}\nonumber\\
                &\quad + \beta^2 \min(t_1,t_2)
            \end{align}
            where $\boldsymbol{\delta}_i:=(1-\delta_i, \delta_i)^\top$ and $L^+\square{(\ul u_1, \ol u_1),(\ul u_2, \ol u_2)}$ represents the $2\times 2$ submatrix of $L^+$ with rows $(\ul u_1,\ol u_1)$ and columns $(\ul u_2,\ol u_2)$.
        \end{proposition}
        Notice that the unique differences with the expression (\ref{eq:covZ}) are the different choices for the temporal kernel $k_T$ and the additional addend $\beta^2 \min(t_1,t_2)$. The latter ensures that the same points at different times have a strictly positive distance, as, combining equations (\ref{eq:distFromKernel}) and (\ref{eq:covZ_Periodic}) for such points $u_1$ and $u_2$, we get $d(u_1,u_2)=\beta^2\abs{t_1-t_2}$.
    
        We conclude this section with a formal assertion regarding the mapping $d$ as being introduced for the case of a periodic graph $\widetilde{\boldsymbol G}$.
        \begin{proposition}
            \label{prop:semiMetricPeriodic}
            $(\widetilde{\boldsymbol G},d)$ is a semi-metric space.
        \end{proposition}

\section{Reproducing kernels} 
    \label{sec:reproducingKernels}
    \subsection{General construction principles}
        Variograms can be composed with certain classes of functions to create reproducing kernels associated with semi-metric spaces. A function $\psi:[0,+\infty)\to \R$ is called \emph{completely monotone} if it is continuous on $[0,+\infty)$, infinitely differentiable on $(0,+\infty)$ and for each $i\in \N$ it holds
        \begin{equation*}
            (-1)^i\,\psi^{(i)}(x)\geq 0,
        \end{equation*}
        where $\psi^{(i)}$ denotes the $i^\text{th}$ derivative of $\psi$ and $\psi^{(0)}:=\psi$. By the celebrated Bernstein's theorem \citep{bernstein_sur_1929}, completely monotone functions are the Laplace transforms of positive and bounded measures. Some examples of parametric families of completely monotone functions are listed in Table \ref{tab:complMonotExamples}.
        
        \begin{table}[t]
            \centering
            \begin{tabular}{|c|c|c|}
                \hline
                Type & $\psi(x)$ & Parameter range\\
                \hline
                Power exponential & $\displaystyle e^{-\beta x^\alpha}$ & $0<\alpha\leq 1$, $\beta>0$ \\
                Mat\'ern & $\displaystyle \frac{2^{1-\alpha}}{\Gamma(\alpha)} (\beta x)^\alpha K_r(\beta x)$ & $0<\alpha\leq \frac{1}{2}$, $\beta>0$\\
                Generalised Cauchy & $\displaystyle (\beta x^\alpha+1)^{-\xi/\alpha}$ & $0<\alpha \leq 1$, $\beta>0$, $\xi>0$\\
                Dagum & $\displaystyle 1 - \round{\frac{\beta x^\alpha}{1+\beta x^\alpha}}^{\xi/\alpha}$ & $0<\alpha\leq 1$, $ 0<\xi \leq 1$, $\beta>0$\\
                \hline
            \end{tabular}
            \caption{Examples of completely monotone functions $0\leq x\mapsto \psi(x)$ such that $\psi(0)=1$. Here, $K_r$ denotes the modified Bessel function of the second kind.}
            \label{tab:complMonotExamples}
        \end{table}
    
        The result below comes straight by using similar arguments as in \cite{anderes2020}, which have been reported in Theorem \ref{theo:covarianceConstruction} in Appendix \ref{app:isotropicKernelsGeneral}.
        
        \begin{proposition} 
            \label{prop:kernel_comp}
            Let $\psi: [0,\infty) \to \R$ be continuous, completely monotonic on the positive real line, and with $\psi(0)< \infty$. Let $d: \widetilde{\boldsymbol G} \times \widetilde{\boldsymbol G} \to \mathbb{R}$ be the mapping defined at (\ref{eq:distFromKernel}). Then, the function 
            \begin{equation*}
                k_{Z}(u_1,u_2) = \psi \round{ d(u_1,u_2) }, \qquad u_1,u_2 \in \widetilde{\boldsymbol G}
            \end{equation*}
            is a strictly positive definite function.
        \end{proposition}
        Proposition \ref{prop:kernel_comp} provides a very easy recipe to build kernels over time evolving graphs, whatever the temporal structure (linear or periodic). Any element from the Table \ref{tab:complMonotExamples} is a good candidate for such a composition. We do not report the corresponding algebraic forms for obvious reasons. Instead, we concentrate on illustrating how these covariance works through two practical examples. We believe that the free parameters and the large number of analytically-tractable completely monotone functions provide a wide range of models that could fit several real-world frameworks. 
    
        \begin{figure}[tp]
            \centering
            \includegraphics[width = 0.7\textwidth]{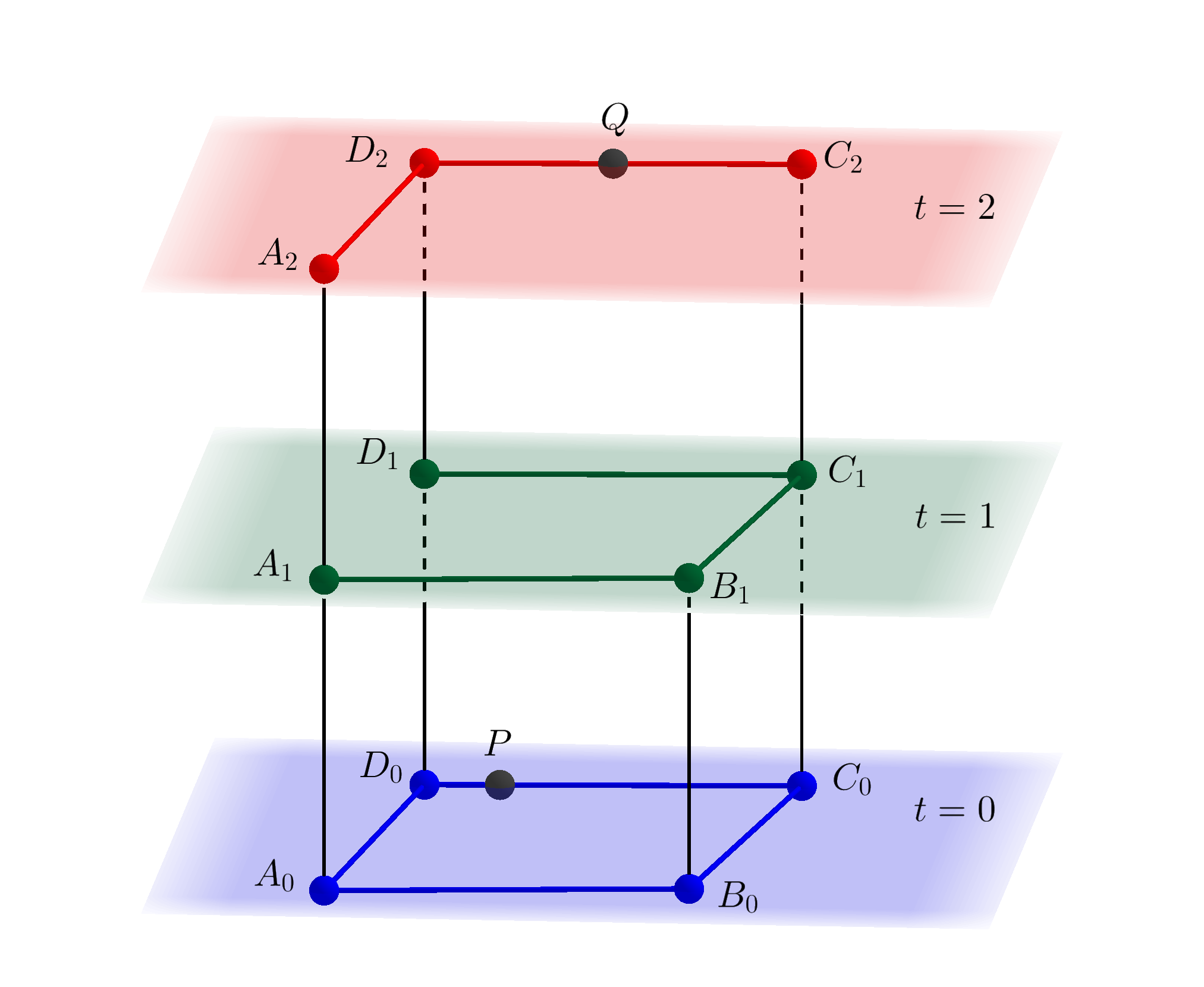}
            \caption{Equivalent simple graph taken as an example for the linear-time case.}
            \label{fig:LayerGraphExample1}
        \end{figure}
    
    \subsection{Linear time}
        We start by considering the graph  in Figure \ref{fig:LayerGraphExample1}. Here, we have $m=3$ time instants. Further, 
        \[
            V=\curly{A_0,B_0,C_0,D_0,A_1,B_1,C_1,D_1,A_2,C_2,D_2}.
        \]
        We focus on the distances as well as the covariances between the points $A_0=\round{A_0,B_0,\delta_{A_0}=0}$, $P:=\round{C_0,D_0,\delta_P=0.8}$ and $Q:=\round{C_2,D_2,\delta_Q=0.5}$. All the spatial edges $E_S$ have weight $1$, whilst the temporal edges have weight $\alpha>0$. Finally, we use $k_T$ as in (\ref{eq:temporalKernel}), with $\lambda$ free parameter. The adjacency matrix follows (here $\cdot$ stands for $0$).
        \begin{footnotesize}
            \begin{equation*}
                \setcounter{MaxMatrixCols}{11}
                \begin{bmatrix}
                    \cdot&1&\cdot&1&\alpha&\cdot&\cdot&\cdot&\cdot&\cdot&\cdot\\
                    1&\cdot&1&\cdot&\cdot&\alpha&\cdot&\cdot&\cdot&\cdot&\cdot\\
                    \cdot&1&\cdot&1&\cdot&\cdot&\alpha&\cdot&\cdot&\cdot&\cdot\\
                    1&\cdot&1&\cdot&\cdot&\cdot&\cdot&\alpha&\cdot&\cdot&\cdot\\
                    \alpha&\cdot&\cdot&\cdot&\cdot&1&\cdot&\cdot&\alpha&\cdot&\cdot\\
                    \cdot&\alpha&\cdot&\cdot&1&\cdot&1&\cdot&\cdot&\cdot&\cdot\\
                    \cdot&\cdot&\alpha&\cdot&\cdot&1&\cdot&1&\cdot&\alpha &\cdot\\
                    \cdot&\cdot&\cdot&\alpha&\cdot&\cdot&1&\cdot&\cdot&\cdot&\alpha\\
                    \cdot&\cdot&\cdot&\cdot&\alpha&\cdot&\cdot&\cdot&\cdot&\cdot&1\\
                    \cdot&\cdot&\cdot&\cdot&\cdot&\cdot&\alpha&\cdot&\cdot&\cdot&1\\
                    \cdot&\cdot&\cdot&\cdot&\cdot&\cdot&\cdot&\alpha&1&1&\cdot\\
                \end{bmatrix}
            \end{equation*}
        \end{footnotesize}
    
        \begin{figure}[p]
            \centering
            \begin{subfigure}[b]{0.48\textwidth}
                \centering
                \begin{tikzpicture}
                    \begin{scriptsize}
                    \begin{axis}[
                        xmin = 0, xmax = 8,
                        ymin = -0.2, ymax = 4,
                        width = \textwidth,
                        height = \textwidth,
                        xlabel = $\alpha$,
                        ]
                        \addplot[color=blue,]
                        coordinates {(0.2,0.7957) (0.3,0.7799) (0.4,0.7664) (0.5,0.7545) (0.6,0.7439) (0.7,0.7343) (0.8,0.7255) (0.9,0.7175) (1,0.7101) (1.1,0.7033) (1.2,0.6969) (1.3,0.6909) (1.4,0.6854) (1.5,0.6801) (1.6,0.6752) (1.7,0.6706) (1.8,0.6662) (1.9,0.662) (2,0.658) (2.1,0.6543) (2.2,0.6507) (2.3,0.6473) (2.4,0.6441) (2.5,0.641) (2.6,0.638) (2.7,0.6351) (2.8,0.6324) (2.9,0.6298) (3,0.6273) (3.1,0.6249) (3.2,0.6225) (3.3,0.6203) (3.4,0.6182) (3.5,0.6161) (3.6,0.6141) (3.7,0.6121) (3.8,0.6103) (3.9,0.6085) (4,0.6067) (4.1,0.605) (4.2,0.6034) (4.3,0.6018) (4.4,0.6002) (4.5,0.5987) (4.6,0.5973) (4.7,0.5959) (4.8,0.5945) (4.9,0.5932) (5,0.5919) (5.1,0.5906) (5.2,0.5894) (5.3,0.5882) (5.4,0.587) (5.5,0.5859) (5.6,0.5848) (5.7,0.5837) (5.8,0.5827) (5.9,0.5817) (6,0.5807) (6.1,0.5797) (6.2,0.5787) (6.3,0.5778) (6.4,0.5769) (6.5,0.576) (6.6,0.5751) (6.7,0.5743) (6.8,0.5735) (6.9,0.5727) (7,0.5719) (7.1,0.5711) (7.2,0.5703) (7.3,0.5696) (7.4,0.5688) (7.5,0.5681) (7.6,0.5674) (7.7,0.5667) (7.8,0.5661) (7.9,0.5654) (8,0.5648)};
                        \addlegendentry{$d(A_0,P)$}
                        \addplot[color=red,]
                        coordinates {(0.2,3.6274) (0.3,2.6445) (0.4,2.1496) (0.5,1.8507) (0.6,1.65) (0.7,1.5058) (0.8,1.3969) (0.9,1.3118) (1,1.2433) (1.1,1.187) (1.2,1.1398) (1.3,1.0997) (1.4,1.0652) (1.5,1.0351) (1.6,1.0088) (1.7,0.9854) (1.8,0.9645) (1.9,0.9458) (2,0.9289) (2.1,0.9136) (2.2,0.8996) (2.3,0.8868) (2.4,0.8751) (2.5,0.8642) (2.6,0.8542) (2.7,0.8448) (2.8,0.8362) (2.9,0.8281) (3,0.8205) (3.1,0.8134) (3.2,0.8068) (3.3,0.8005) (3.4,0.7946) (3.5,0.789) (3.6,0.7837) (3.7,0.7787) (3.8,0.7739) (3.9,0.7694) (4,0.7651) (4.1,0.7611) (4.2,0.7571) (4.3,0.7534) (4.4,0.7499) (4.5,0.7465) (4.6,0.7432) (4.7,0.7401) (4.8,0.7371) (4.9,0.7342) (5,0.7314) (5.1,0.7288) (5.2,0.7262) (5.3,0.7237) (5.4,0.7214) (5.5,0.7191) (5.6,0.7168) (5.7,0.7147) (5.8,0.7126) (5.9,0.7106) (6,0.7087) (6.1,0.7068) (6.2,0.705) (6.3,0.7033) (6.4,0.7016) (6.5,0.6999) (6.6,0.6983) (6.7,0.6967) (6.8,0.6952) (6.9,0.6938) (7,0.6923) (7.1,0.691) (7.2,0.6896) (7.3,0.6883) (7.4,0.687) (7.5,0.6858) (7.6,0.6845) (7.7,0.6834) (7.8,0.6822) (7.9,0.6811) (8,0.68)};
                        \addlegendentry{$d(A_0,Q)$}
                    \end{axis}
                    \end{scriptsize}
                \end{tikzpicture}
            \end{subfigure}
            \begin{subfigure}[b]{0.48\textwidth}
                \centering
                \begin{tikzpicture}
                    \begin{scriptsize}
                    \begin{axis}[
                        xmin = 0, xmax = 8,
                        ymin = -0.2, ymax = 4,
                        width = \textwidth,
                        height = \textwidth,
                        xlabel = $\alpha$,
                        ]
                        \addplot[color=blue,]
                        coordinates {(0.2,3.6478) (0.3,2.6561) (0.4,2.1534) (0.5,1.8473) (0.6,1.6402) (0.7,1.49) (0.8,1.3757) (0.9,1.2856) (1,1.2125) (1.1,1.1518) (1.2,1.1007) (1.3,1.0569) (1.4,1.0189) (1.5,0.9856) (1.6,0.9561) (1.7,0.9299) (1.8,0.9063) (1.9,0.8851) (2,0.8657) (2.1,0.8481) (2.2,0.8319) (2.3,0.817) (2.4,0.8033) (2.5,0.7905) (2.6,0.7787) (2.7,0.7677) (2.8,0.7573) (2.9,0.7477) (3,0.7386) (3.1,0.73) (3.2,0.722) (3.3,0.7143) (3.4,0.7071) (3.5,0.7003) (3.6,0.6938) (3.7,0.6877) (3.8,0.6818) (3.9,0.6762) (4,0.6709) (4.1,0.6658) (4.2,0.6609) (4.3,0.6562) (4.4,0.6517) (4.5,0.6475) (4.6,0.6433) (4.7,0.6394) (4.8,0.6356) (4.9,0.6319) (5,0.6284) (5.1,0.625) (5.2,0.6217) (5.3,0.6185) (5.4,0.6155) (5.5,0.6125) (5.6,0.6096) (5.7,0.6069) (5.8,0.6042) (5.9,0.6016) (6,0.5991) (6.1,0.5966) (6.2,0.5943) (6.3,0.592) (6.4,0.5897) (6.5,0.5876) (6.6,0.5855) (6.7,0.5834) (6.8,0.5814) (6.9,0.5795) (7,0.5776) (7.1,0.5758) (7.2,0.574) (7.3,0.5722) (7.4,0.5705) (7.5,0.5689) (7.6,0.5672) (7.7,0.5657) (7.8,0.5641) (7.9,0.5626) (8,0.5611)};
                        \addlegendentry{$\lambda = 0$}
                        \addplot[color=red,]
                        coordinates {(0.2,3.5758) (0.3,2.5841) (0.4,2.0814) (0.5,1.7753) (0.6,1.5682) (0.7,1.418) (0.8,1.3037) (0.9,1.2136) (1,1.1405) (1.1,1.0798) (1.2,1.0287) (1.3,0.9849) (1.4,0.9469) (1.5,0.9136) (1.6,0.8841) (1.7,0.8579) (1.8,0.8343) (1.9,0.8131) (2,0.7937) (2.1,0.7761) (2.2,0.7599) (2.3,0.745) (2.4,0.7313) (2.5,0.7185) (2.6,0.7067) (2.7,0.6957) (2.8,0.6853) (2.9,0.6757) (3,0.6666) (3.1,0.658) (3.2,0.65) (3.3,0.6423) (3.4,0.6351) (3.5,0.6283) (3.6,0.6218) (3.7,0.6157) (3.8,0.6098) (3.9,0.6042) (4,0.5989) (4.1,0.5938) (4.2,0.5889) (4.3,0.5842) (4.4,0.5797) (4.5,0.5755) (4.6,0.5713) (4.7,0.5674) (4.8,0.5636) (4.9,0.5599) (5,0.5564) (5.1,0.553) (5.2,0.5497) (5.3,0.5465) (5.4,0.5435) (5.5,0.5405) (5.6,0.5376) (5.7,0.5349) (5.8,0.5322) (5.9,0.5296) (6,0.5271) (6.1,0.5246) (6.2,0.5223) (6.3,0.52) (6.4,0.5177) (6.5,0.5156) (6.6,0.5135) (6.7,0.5114) (6.8,0.5094) (6.9,0.5075) (7,0.5056) (7.1,0.5038) (7.2,0.502) (7.3,0.5002) (7.4,0.4985) (7.5,0.4969) (7.6,0.4952) (7.7,0.4937) (7.8,0.4921) (7.9,0.4906) (8,0.4891)};
                        \addlegendentry{$\lambda = 0.6$}
                        \addplot[color=green,]
                        coordinates {(0.2,3.4858) (0.3,2.4941) (0.4,1.9914) (0.5,1.6853) (0.6,1.4782) (0.7,1.328) (0.8,1.2137) (0.9,1.1236) (1,1.0505) (1.1,0.9898) (1.2,0.9387) (1.3,0.8949) (1.4,0.8569) (1.5,0.8236) (1.6,0.7941) (1.7,0.7679) (1.8,0.7443) (1.9,0.7231) (2,0.7037) (2.1,0.6861) (2.2,0.6699) (2.3,0.655) (2.4,0.6413) (2.5,0.6285) (2.6,0.6167) (2.7,0.6057) (2.8,0.5953) (2.9,0.5857) (3,0.5766) (3.1,0.568) (3.2,0.56) (3.3,0.5523) (3.4,0.5451) (3.5,0.5383) (3.6,0.5318) (3.7,0.5257) (3.8,0.5198) (3.9,0.5142) (4,0.5089) (4.1,0.5038) (4.2,0.4989) (4.3,0.4942) (4.4,0.4897) (4.5,0.4855) (4.6,0.4813) (4.7,0.4774) (4.8,0.4736) (4.9,0.4699) (5,0.4664) (5.1,0.463) (5.2,0.4597) (5.3,0.4565) (5.4,0.4535) (5.5,0.4505) (5.6,0.4476) (5.7,0.4449) (5.8,0.4422) (5.9,0.4396) (6,0.4371) (6.1,0.4346) (6.2,0.4323) (6.3,0.43) (6.4,0.4277) (6.5,0.4256) (6.6,0.4235) (6.7,0.4214) (6.8,0.4194) (6.9,0.4175) (7,0.4156) (7.1,0.4138) (7.2,0.412) (7.3,0.4102) (7.4,0.4085) (7.5,0.4069) (7.6,0.4052) (7.7,0.4037) (7.8,0.4021) (7.9,0.4006) (8,0.3991)};
                        \addlegendentry{$\lambda = 0.9$}
                    \end{axis}
                    \end{scriptsize}
                \end{tikzpicture}
            \end{subfigure}
            \caption{Distances between the points $A_0$, $P$ and $Q$. Notice that while $d(A_0,P)$ and $d(A_0,Q)$ (left) do not depend on $\lambda$, the distance $d(P,Q)$ (right) decreases as $\lambda$ increases.}
            \label{fig:exampleDists6_LinearTime}
        \end{figure}
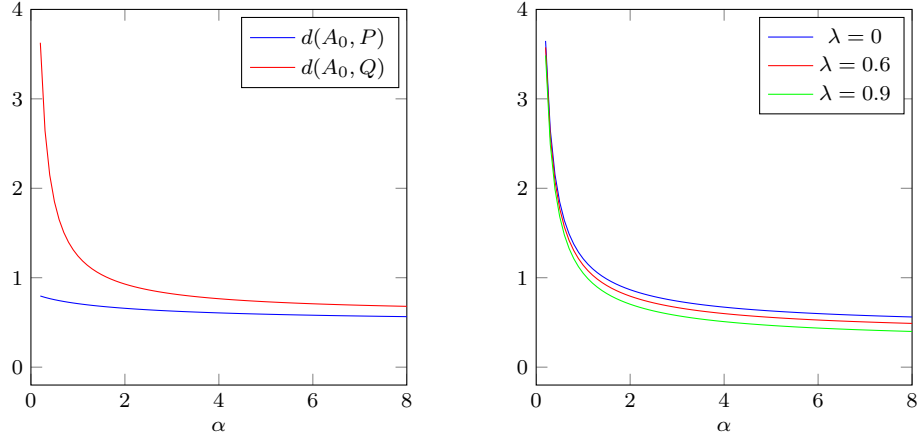
    
        \begin{figure}[p]
            \centering
            \begin{subfigure}[b]{0.48\textwidth}
                \centering
                \begin{tikzpicture}
                    \begin{scriptsize}
                    \begin{axis}[
                        xmin = 0, xmax = 8,
                        ymin = 0, ymax = 1,
                        width = \textwidth,
                        height = \textwidth,
                        xlabel = $\alpha$,
                        ]
                        \addplot[color=blue,]
                        coordinates {(0.2,0.4513) (0.3,0.4584) (0.4,0.4647) (0.5,0.4702) (0.6,0.4753) (0.7,0.4799) (0.8,0.4841) (0.9,0.488) (1,0.4916) (1.1,0.495) (1.2,0.4981) (1.3,0.5011) (1.4,0.5039) (1.5,0.5066) (1.6,0.5091) (1.7,0.5114) (1.8,0.5137) (1.9,0.5158) (2,0.5179) (2.1,0.5198) (2.2,0.5217) (2.3,0.5235) (2.4,0.5252) (2.5,0.5268) (2.6,0.5284) (2.7,0.5299) (2.8,0.5313) (2.9,0.5327) (3,0.534) (3.1,0.5353) (3.2,0.5366) (3.3,0.5378) (3.4,0.5389) (3.5,0.5401) (3.6,0.5411) (3.7,0.5422) (3.8,0.5432) (3.9,0.5442) (4,0.5451) (4.1,0.5461) (4.2,0.547) (4.3,0.5478) (4.4,0.5487) (4.5,0.5495) (4.6,0.5503) (4.7,0.5511) (4.8,0.5518) (4.9,0.5526) (5,0.5533) (5.1,0.554) (5.2,0.5547) (5.3,0.5553) (5.4,0.556) (5.5,0.5566) (5.6,0.5572) (5.7,0.5578) (5.8,0.5584) (5.9,0.559) (6,0.5595) (6.1,0.5601) (6.2,0.5606) (6.3,0.5611) (6.4,0.5616) (6.5,0.5621) (6.6,0.5626) (6.7,0.5631) (6.8,0.5636) (6.9,0.564) (7,0.5645) (7.1,0.5649) (7.2,0.5653) (7.3,0.5658) (7.4,0.5662) (7.5,0.5666) (7.6,0.567) (7.7,0.5674) (7.8,0.5677) (7.9,0.5681) (8,0.5685)};
                        \addlegendentry{$\Cov{A_0}{P}$}
                        \addplot[color=red,]
                        coordinates {(0.2,0.0266) (0.3,0.071) (0.4,0.1165) (0.5,0.1571) (0.6,0.192) (0.7,0.2218) (0.8,0.2474) (0.9,0.2693) (1,0.2884) (1.1,0.3051) (1.2,0.3199) (1.3,0.333) (1.4,0.3447) (1.5,0.3552) (1.6,0.3647) (1.7,0.3733) (1.8,0.3812) (1.9,0.3884) (2,0.395) (2.1,0.4011) (2.2,0.4067) (2.3,0.412) (2.4,0.4168) (2.5,0.4214) (2.6,0.4256) (2.7,0.4296) (2.8,0.4334) (2.9,0.4369) (3,0.4402) (3.1,0.4433) (3.2,0.4463) (3.3,0.4491) (3.4,0.4518) (3.5,0.4543) (3.6,0.4567) (3.7,0.459) (3.8,0.4612) (3.9,0.4633) (4,0.4653) (4.1,0.4672) (4.2,0.469) (4.3,0.4708) (4.4,0.4724) (4.5,0.474) (4.6,0.4756) (4.7,0.4771) (4.8,0.4785) (4.9,0.4799) (5,0.4812) (5.1,0.4825) (5.2,0.4837) (5.3,0.4849) (5.4,0.4861) (5.5,0.4872) (5.6,0.4883) (5.7,0.4893) (5.8,0.4903) (5.9,0.4913) (6,0.4923) (6.1,0.4932) (6.2,0.4941) (6.3,0.495) (6.4,0.4958) (6.5,0.4966) (6.6,0.4974) (6.7,0.4982) (6.8,0.499) (6.9,0.4997) (7,0.5004) (7.1,0.5011) (7.2,0.5018) (7.3,0.5024) (7.4,0.5031) (7.5,0.5037) (7.6,0.5043) (7.7,0.5049) (7.8,0.5055) (7.9,0.5061) (8,0.5066)};
                        \addlegendentry{$\Cov{A_0}{Q}$}
                        \addplot[color=green,]
                        coordinates {(0.2,0.028) (0.3,0.0755) (0.4,0.1248) (0.5,0.1694) (0.6,0.2084) (0.7,0.2422) (0.8,0.2715) (0.9,0.2971) (1,0.3197) (1.1,0.3396) (1.2,0.3575) (1.3,0.3735) (1.4,0.3879) (1.5,0.4011) (1.6,0.4131) (1.7,0.4241) (1.8,0.4342) (1.9,0.4435) (2,0.4522) (2.1,0.4602) (2.2,0.4677) (2.3,0.4747) (2.4,0.4813) (2.5,0.4875) (2.6,0.4933) (2.7,0.4987) (2.8,0.5039) (2.9,0.5088) (3,0.5135) (3.1,0.5179) (3.2,0.5221) (3.3,0.5261) (3.4,0.5299) (3.5,0.5335) (3.6,0.537) (3.7,0.5403) (3.8,0.5435) (3.9,0.5465) (4,0.5494) (4.1,0.5522) (4.2,0.5549) (4.3,0.5575) (4.4,0.56) (4.5,0.5624) (4.6,0.5648) (4.7,0.567) (4.8,0.5692) (4.9,0.5713) (5,0.5733) (5.1,0.5752) (5.2,0.5771) (5.3,0.579) (5.4,0.5807) (5.5,0.5825) (5.6,0.5841) (5.7,0.5857) (5.8,0.5873) (5.9,0.5888) (6,0.5903) (6.1,0.5918) (6.2,0.5932) (6.3,0.5945) (6.4,0.5959) (6.5,0.5972) (6.6,0.5984) (6.7,0.5996) (6.8,0.6008) (6.9,0.602) (7,0.6031) (7.1,0.6043) (7.2,0.6053) (7.3,0.6064) (7.4,0.6074) (7.5,0.6084) (7.6,0.6094) (7.7,0.6104) (7.8,0.6113) (7.9,0.6123) (8,0.6132)};
                        \addlegendentry{$\Cov{P}{Q}\, (\lambda = 0.6)$}
                    \end{axis}
                    \end{scriptsize}
                \end{tikzpicture}
            \end{subfigure}
            \begin{subfigure}[b]{0.48\textwidth}
                \centering
                \begin{tikzpicture}
                    \begin{scriptsize}
                    \begin{axis}[
                        xmin = 0, xmax = 8,
                        ymin = 0, ymax = 1,
                        width = \textwidth,
                        height = \textwidth,
                        xlabel = $\alpha$,
                        ]
                        \addplot[color=blue,]
                        coordinates {(0.2,0.4482) (0.3,0.4518) (0.4,0.4549) (0.5,0.4578) (0.6,0.4603) (0.7,0.4627) (0.8,0.4649) (0.9,0.4669) (1,0.4688) (1.1,0.4706) (1.2,0.4722) (1.3,0.4738) (1.4,0.4753) (1.5,0.4767) (1.6,0.478) (1.7,0.4793) (1.8,0.4805) (1.9,0.4817) (2,0.4828) (2.1,0.4839) (2.2,0.4849) (2.3,0.4858) (2.4,0.4868) (2.5,0.4877) (2.6,0.4885) (2.7,0.4894) (2.8,0.4902) (2.9,0.4909) (3,0.4917) (3.1,0.4924) (3.2,0.4931) (3.3,0.4938) (3.4,0.4944) (3.5,0.4951) (3.6,0.4957) (3.7,0.4963) (3.8,0.4968) (3.9,0.4974) (4,0.4979) (4.1,0.4984) (4.2,0.499) (4.3,0.4994) (4.4,0.4999) (4.5,0.5004) (4.6,0.5009) (4.7,0.5013) (4.8,0.5017) (4.9,0.5021) (5,0.5026) (5.1,0.503) (5.2,0.5034) (5.3,0.5037) (5.4,0.5041) (5.5,0.5045) (5.6,0.5048) (5.7,0.5052) (5.8,0.5055) (5.9,0.5058) (6,0.5062) (6.1,0.5065) (6.2,0.5068) (6.3,0.5071) (6.4,0.5074) (6.5,0.5077) (6.6,0.508) (6.7,0.5082) (6.8,0.5085) (6.9,0.5088) (7,0.509) (7.1,0.5093) (7.2,0.5095) (7.3,0.5098) (7.4,0.51) (7.5,0.5103) (7.6,0.5105) (7.7,0.5107) (7.8,0.511) (7.9,0.5112) (8,0.5114)};
                        \addlegendentry{$\Cov{A_0}{P}$}
                        \addplot[color=red,]
                        coordinates {(0.2,0.2286) (0.3,0.2652) (0.4,0.2918) (0.5,0.3123) (0.6,0.3288) (0.7,0.3424) (0.8,0.3539) (0.9,0.3637) (1,0.3723) (1.1,0.3797) (1.2,0.3864) (1.3,0.3923) (1.4,0.3976) (1.5,0.4024) (1.6,0.4068) (1.7,0.4108) (1.8,0.4144) (1.9,0.4178) (2,0.4209) (2.1,0.4238) (2.2,0.4265) (2.3,0.429) (2.4,0.4313) (2.5,0.4335) (2.6,0.4356) (2.7,0.4375) (2.8,0.4393) (2.9,0.4411) (3,0.4427) (3.1,0.4442) (3.2,0.4457) (3.3,0.4471) (3.4,0.4484) (3.5,0.4497) (3.6,0.4509) (3.7,0.4521) (3.8,0.4532) (3.9,0.4542) (4,0.4552) (4.1,0.4562) (4.2,0.4571) (4.3,0.458) (4.4,0.4589) (4.5,0.4597) (4.6,0.4605) (4.7,0.4612) (4.8,0.462) (4.9,0.4627) (5,0.4634) (5.1,0.464) (5.2,0.4647) (5.3,0.4653) (5.4,0.4659) (5.5,0.4665) (5.6,0.4671) (5.7,0.4676) (5.8,0.4681) (5.9,0.4686) (6,0.4691) (6.1,0.4696) (6.2,0.4701) (6.3,0.4706) (6.4,0.471) (6.5,0.4714) (6.6,0.4718) (6.7,0.4723) (6.8,0.4727) (6.9,0.473) (7,0.4734) (7.1,0.4738) (7.2,0.4742) (7.3,0.4745) (7.4,0.4748) (7.5,0.4752) (7.6,0.4755) (7.7,0.4758) (7.8,0.4761) (7.9,0.4764) (8,0.4767)};
                        \addlegendentry{$\Cov{A_0}{Q}$}
                        \addplot[color=green,]
                        coordinates {(0.2,0.2301) (0.3,0.268) (0.4,0.2961) (0.5,0.3182) (0.6,0.3363) (0.7,0.3516) (0.8,0.3647) (0.9,0.3761) (1,0.3863) (1.1,0.3953) (1.2,0.4035) (1.3,0.4108) (1.4,0.4176) (1.5,0.4238) (1.6,0.4295) (1.7,0.4348) (1.8,0.4397) (1.9,0.4443) (2,0.4486) (2.1,0.4527) (2.2,0.4565) (2.3,0.46) (2.4,0.4634) (2.5,0.4666) (2.6,0.4697) (2.7,0.4725) (2.8,0.4753) (2.9,0.4779) (3,0.4804) (3.1,0.4828) (3.2,0.4851) (3.3,0.4873) (3.4,0.4894) (3.5,0.4914) (3.6,0.4933) (3.7,0.4952) (3.8,0.497) (3.9,0.4987) (4,0.5004) (4.1,0.502) (4.2,0.5035) (4.3,0.505) (4.4,0.5065) (4.5,0.5079) (4.6,0.5092) (4.7,0.5105) (4.8,0.5118) (4.9,0.513) (5,0.5142) (5.1,0.5154) (5.2,0.5165) (5.3,0.5176) (5.4,0.5187) (5.5,0.5197) (5.6,0.5207) (5.7,0.5217) (5.8,0.5226) (5.9,0.5236) (6,0.5245) (6.1,0.5254) (6.2,0.5262) (6.3,0.5271) (6.4,0.5279) (6.5,0.5287) (6.6,0.5295) (6.7,0.5302) (6.8,0.531) (6.9,0.5317) (7,0.5324) (7.1,0.5331) (7.2,0.5338) (7.3,0.5344) (7.4,0.5351) (7.5,0.5357) (7.6,0.5364) (7.7,0.537) (7.8,0.5376) (7.9,0.5381) (8,0.5387)};
                        \addlegendentry{$\Cov{P}{Q}\, (\lambda = 0.6)$}
                    \end{axis}
                    \end{scriptsize}
                \end{tikzpicture}
            \end{subfigure}
            \caption{Generated covariances between the points $A_0$, $P$ and $Q$. Left: exponential kernel with parameters $(\alpha=1,\beta=1)$ (see Table \ref{tab:complMonotExamples}). Right: generalised Cauchy kernel with parameters $(\alpha = 1,\beta =5,\xi=0.5)$ (see Table \ref{tab:complMonotExamples}).}
            \label{fig:exampleDists6_LinearTimeCov}
        \end{figure}
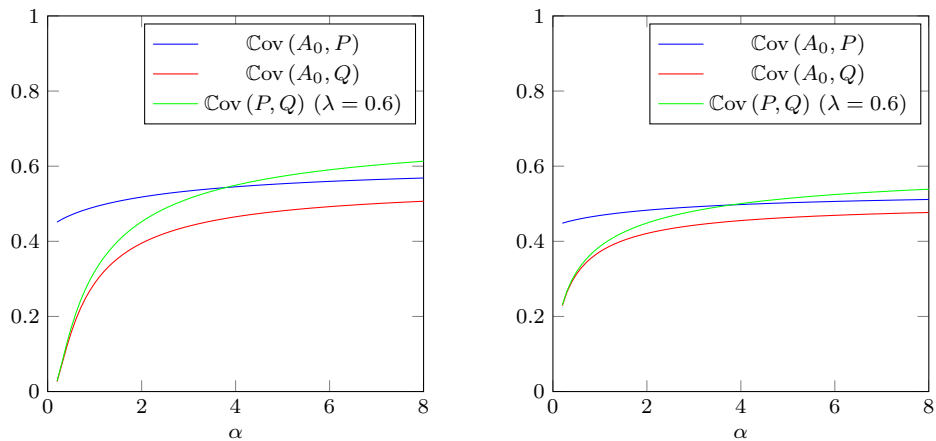
        
        Figure \ref{fig:exampleDists6_LinearTime} clearly shows the effect the temporal edge parameter $\alpha$ plays on the distances: while it has a considerable impact on the distances $d(A_0,Q)$ and $d(P,Q)$, it shows a negligible effect on $d(A_0,P)$. This is reasonable given the graph structure: $A_0$ and $P$ belong to the same layer ($t=0$) and they are connected from both the paths $A_0,D_0,P$ and $A_0,B_0,C_0,P$, which completely lie on $t=0$ (and therefore they do not change with $\alpha$). On the other hand, $Q$ can be connected to $A_0$ and $P$ only via paths that include temporal edges. As a consequence, if $\alpha \to 0^+$, both $d(A_0,Q)$ and $d(P,Q)$ will go to infinity. \par
        The plot on the right of Figure \ref{fig:exampleDists6_LinearTime} shows the effect of the correlation parameter $\lambda$ as well. Whilst it does not influence the distances concerning $A_0$ (since it is a vertex), as it increase, it reduces the distances between $P$ and $Q$. Clearly, the effect is more significant for large values of the parameter $\alpha$. Indeed, when $\alpha$ is small, the distances between nodes at different time instants are large. As a consequence, the second line of equation (\ref{eq:covZ}) becomes negligible when compared to the first one.\par
        Figure \ref{fig:exampleDists6_LinearTimeCov} shows the resulting effect of the parameter $\alpha$ on the correlations between $A_0$, $P$ and $Q$ generated by the composition of two completely monotone functions taken from Table \ref{tab:complMonotExamples} and the distances shown in Figure \ref{fig:exampleDists6_LinearTime}.
        
    \subsection{Circular time}
        \begin{figure}[tp]
            \centering
            \resizebox{0.4\textwidth}{!}{
                \begin{tikzpicture}
                    \node (A) at (0,0) {$A_0$};
                    \node (B) at (4,0) {$B_0$};
                    \node (C) at (0,1) {$A_1$};
                    \node (D) at (4,1) {$B_1$};
                    \node (E) at (0,2) {$A_2$};
                    \node (F) at (4,2) {$B_2$};
                    \node (G) at (0,3) {$A_3$};
                    \node (H) at (4,3) {$B_3$};
                    \node (I) at (0,4) {$A_4$};
                    \node (J) at (4,4) {$B_4$};
                    \node (K) at (0,5) {$A_5$};
                    \node (L) at (4,5) {$B_5$};
                    \node (M) at (0,6) {$A_6$};
                    \node (N) at (4,6) {$B_6$};
                    \node (O) at (0,7) {$A_7$};
                    \node (P) at (4,7) {$B_7$};
                    
                    \node (t0) at (6,0) {$\tau=0$};
                    \node (t1) at (6,1) {$\tau=1$};
                    \node (t2) at (6,2) {$\tau=2$};
                    \node (t1) at (6,3) {$\tau=3$};
                    \node (t2) at (6,4) {$\tau=4$};
                    \node (t2) at (6,5) {$\tau=5$};
                    \node (t2) at (6,6) {$\tau=6$};
                    \node (t2) at (6,7) {$\tau=7$};
    
                    \filldraw (2,0) circle[radius=1.5pt];
                    \filldraw (2,1) circle[radius=1.5pt];
                    \filldraw (2,2) circle[radius=1.5pt];
                    \filldraw (2,3) circle[radius=1.5pt];
                    \filldraw (2,4) circle[radius=1.5pt];
                    \filldraw (2,5) circle[radius=1.5pt];
                    \filldraw (2,6) circle[radius=1.5pt];
                    \filldraw (2,7) circle[radius=1.5pt];
                    
                    \node[above=3pt, outer sep=0pt] at (2,0) {$P_0$};
                    \node[above=3pt, outer sep=0pt] at (2,1) {$P_1$};
                    \node[above=3pt, outer sep=0pt] at (2,2) {$P_2$};
                    \node[above=3pt, outer sep=0pt] at (2,3) {$P_3$};
                    \node[above=3pt, outer sep=0pt] at (2,4) {$P_4$};
                    \node[above=3pt, outer sep=0pt] at (2,5) {$P_5$};
                    \node[above=3pt, outer sep=0pt] at (2,6) {$P_6$};
                    \node[above=3pt, outer sep=0pt] at (2,7) {$P_7$};
                
                    \path [-] (A) edge (B);
                    \path [-] (C) edge (D);
                    \path [-] (E) edge (F);
                    \path [-] (G) edge (H);
                    \path [-] (I) edge (J);
                    \path [-] (K) edge (L);
                    \path [-] (M) edge (N);
                    \path [-] (O) edge (P);
                    
                    \path [-] (A) edge (C);
                    \path [-] (B) edge (D);
                    \path [-] (C) edge (E);
                    \path [-] (D) edge (F);
                    \path [-] (E) edge (G);
                    \path [-] (F) edge (H);
                    \path [-] (G) edge (I);
                    \path [-] (H) edge (J);
                    \path [-] (I) edge (K);
                    \path [-] (J) edge (L);
                    \path [-] (K) edge (M);
                    \path [-] (L) edge (N);
                    \path [-] (M) edge (O);
                    \path [-] (N) edge (P);
                    \path [style={bend left=160}] (A) edge (O);
                    \path [style={bend right=160}] (B) edge (P);
                \end{tikzpicture}
            }
            \caption{Equivalent simple graph taken as an example for the circular-time case.}
            \label{fig:LayerGraphExample2}
        \end{figure}
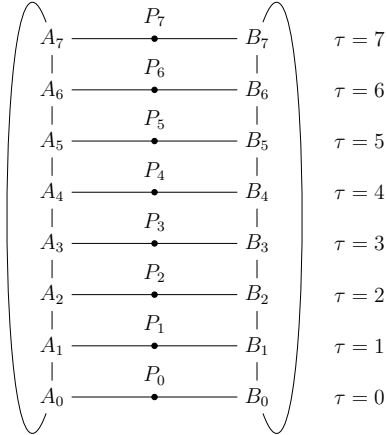
        We are going to analyse the time-evolving periodic graph depicted in Figure \ref{fig:LayerGraphExample2}. In this case, we have $m=8$ and 
        \begin{equation*}
            V=\curly{A_0,B_0,A_1,B_1,\dots,A_7,B_7}.
        \end{equation*}
        We will compare the distances and the covariances between the points $P_0:=\round{0, A_0, B_0, \delta_{P_0}=0.5}$ and $P_t:=\round{t,A_\tau,B_\tau,\delta_{P_t}=0.5}$, where $t\in \N$ and $\tau \equiv t \pmod{m}$. Here, all the spatial edges have weight $1$, whilst the temporal ones have weight $\alpha>0$. We use the temporal kernel $k_T$ as described in Subsection \ref{ssec:circTime_lifespanFull}, with $\rho$ and $\beta$ free parameters. 
        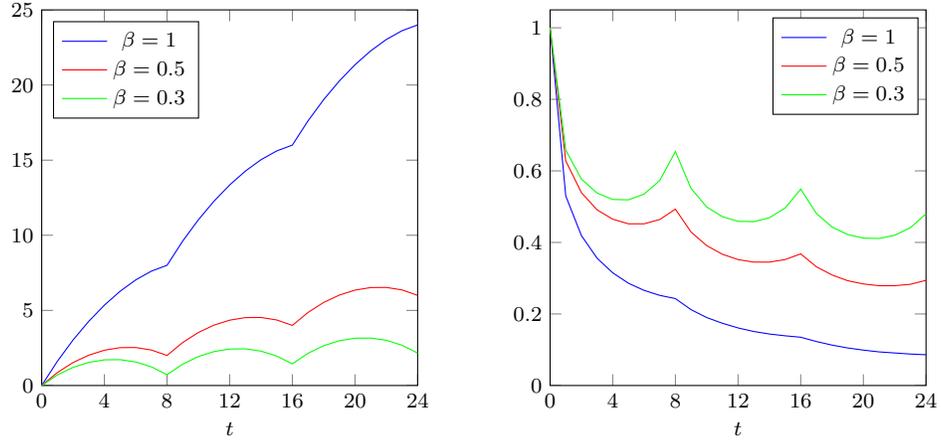
\begin{figure}[p]
            \centering
            \begin{subfigure}[b]{0.48\textwidth}
                \centering
                \begin{tikzpicture}
                    \begin{scriptsize}
                    \begin{axis}[
                        xmin = 0, xmax = 24,
                        ymin = 0, ymax = 25,
                        width = \textwidth,
                        height = \textwidth,
                        xlabel = $t$,
                        legend pos = north west,
                        xtick = {0, 4, 8, 12, 16, 20, 24},
                        ]
                        \addplot[color=blue,]
                        coordinates {(0,0) (1,1.613) (2,3.029) (3,4.27) (4,5.349) (5,6.27) (6,7.029) (7,7.613) (8,8) (9,9.613) (10,11.029) (11,12.27) (12,13.349) (13,14.27) (14,15.029) (15,15.613) (16,16) (17,17.613) (18,19.029) (19,20.27) (20,21.349) (21,22.27) (22,23.029) (23,23.613) (24,24)};
                        \addlegendentry{$\beta = 1$}
                        \addplot[color=red,]
                        coordinates {(0,0) (1,0.863) (2,1.529) (3,2.02) (4,2.349) (5,2.52) (6,2.529) (7,2.363) (8,2) (9,2.863) (10,3.529) (11,4.02) (12,4.349) (13,4.52) (14,4.529) (15,4.363) (16,4) (17,4.863) (18,5.529) (19,6.02) (20,6.349) (21,6.52) (22,6.529) (23,6.363) (24,6)};
                        \addlegendentry{$\beta = 0.5$}
                        \addplot[color=green,]
                        coordinates {(0,0) (1,0.703) (2,1.209) (3,1.54) (4,1.709) (5,1.72) (6,1.569) (7,1.243) (8,0.72) (9,1.423) (10,1.929) (11,2.26) (12,2.429) (13,2.44) (14,2.289) (15,1.963) (16,1.44) (17,2.143) (18,2.649) (19,2.98) (20,3.149) (21,3.16) (22,3.009) (23,2.683) (24,2.16)};
                        \addlegendentry{$\beta = 0.3$}
                    \end{axis}
                    \end{scriptsize}
                \end{tikzpicture}
            \end{subfigure}
            \begin{subfigure}[b]{0.48\textwidth}
                \centering
                \begin{tikzpicture}
                    \begin{scriptsize}
                    \begin{axis}[
                        xmin = 0, xmax = 24,
                        ymin = 0, ymax = 1.05,
                        width = \textwidth,
                        height = \textwidth,
                        xlabel = $t$,
                        xtick = {0, 4, 8, 12, 16, 20, 24},
                        ]
                        \addplot[color=blue,]
                        coordinates {(0,1) (1,0.53) (2,0.419) (3,0.356) (4,0.315) (5,0.286) (6,0.266) (7,0.252) (8,0.243) (9,0.212) (10,0.19) (11,0.174) (12,0.161) (13,0.151) (14,0.144) (15,0.139) (16,0.135) (17,0.123) (18,0.113) (19,0.105) (20,0.099) (21,0.094) (22,0.091) (23,0.088) (24,0.086)};
                        \addlegendentry{$\beta = 1$}
                        \addplot[color=red,]
                        coordinates {(0,1) (1,0.628) (2,0.539) (3,0.491) (4,0.465) (5,0.452) (6,0.452) (7,0.464) (8,0.493) (9,0.429) (10,0.391) (11,0.367) (12,0.352) (13,0.345) (14,0.345) (15,0.352) (16,0.368) (17,0.332) (18,0.309) (19,0.293) (20,0.284) (21,0.279) (22,0.279) (23,0.283) (24,0.294)};
                        \addlegendentry{$\beta = 0.5$}
                        \addplot[color=green,]
                        coordinates {(0,1) (1,0.658) (2,0.577) (3,0.538) (4,0.52) (5,0.519) (6,0.535) (7,0.573) (8,0.654) (9,0.551) (10,0.499) (11,0.472) (12,0.459) (13,0.458) (14,0.469) (15,0.496) (16,0.549) (17,0.481) (18,0.443) (19,0.422) (20,0.412) (21,0.411) (22,0.42) (23,0.441) (24,0.48)};
                        \addlegendentry{$\beta = 0.3$}
                    \end{axis}
                    \end{scriptsize}
                \end{tikzpicture}
            \end{subfigure}
            \caption{Distances (left) and covariances (right) for the graph in Figure \ref{fig:LayerGraphExample2} between the points $P_0$ and $P_t$ for $\rho = 0.45$ and $\alpha = 1$. Covariances have been generated via the exponential kernel with parameters $\alpha=0.5$ and $\beta = 0.5$ (see Table \ref{tab:complMonotExamples}).}
            \label{fig:exampleDists6_CircTimeDists}
        \end{figure}
        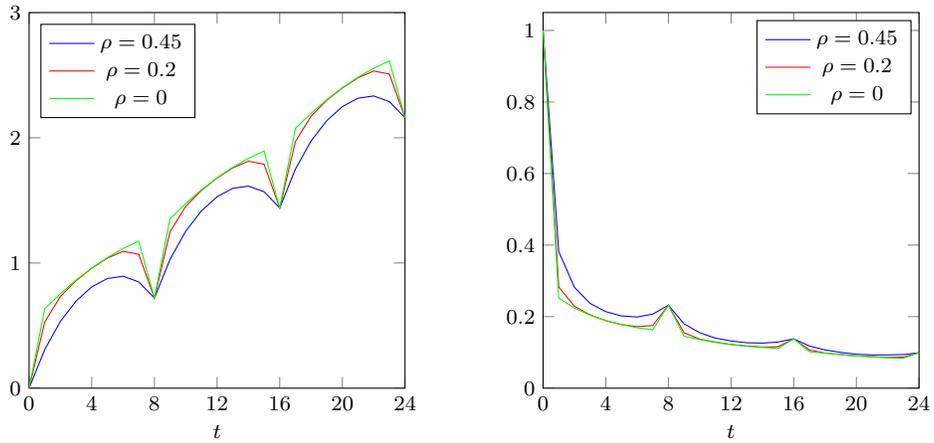
\begin{figure}[p]
            \centering
            \begin{subfigure}[b]{0.48\textwidth}
                \centering
                \begin{tikzpicture}
                    \begin{scriptsize}
                    \begin{axis}[
                        xmin = 0, xmax = 24,
                        ymin = 0, ymax = 3,
                        width = \textwidth,
                        height = \textwidth,
                        xlabel = $t$,
                        legend pos = north west,
                        xtick = {0, 4, 8, 12, 16, 20, 24},
                        ]
                        \addplot[color=blue,]
                        coordinates {(0,0) (1,0.309) (2,0.534) (3,0.696) (4,0.809) (5,0.876) (6,0.894) (7,0.849) (8,0.72) (9,1.029) (10,1.254) (11,1.416) (12,1.529) (13,1.596) (14,1.614) (15,1.569) (16,1.44) (17,1.749) (18,1.974) (19,2.136) (20,2.249) (21,2.316) (22,2.334) (23,2.289) (24,2.16)};
                        \addlegendentry{$\rho = 0.45$}
                        \addplot[color=red,]
                        coordinates {(0,0) (1,0.529) (2,0.733) (3,0.859) (4,0.958) (5,1.039) (6,1.093) (7,1.069) (8,0.72) (9,1.249) (10,1.453) (11,1.579) (12,1.678) (13,1.759) (14,1.813) (15,1.789) (16,1.44) (17,1.969) (18,2.173) (19,2.299) (20,2.398) (21,2.479) (22,2.533) (23,2.509) (24,2.16)};
                        \addlegendentry{$\rho = 0.2$}
                        \addplot[color=green,]
                        coordinates {(0,0) (1,0.634) (2,0.755) (3,0.864) (4,0.96) (5,1.044) (6,1.115) (7,1.174) (8,0.72) (9,1.354) (10,1.475) (11,1.584) (12,1.68) (13,1.764) (14,1.835) (15,1.894) (16,1.44) (17,2.074) (18,2.195) (19,2.304) (20,2.4) (21,2.484) (22,2.555) (23,2.614) (24,2.16)};
                        \addlegendentry{$\rho = 0$}
                    \end{axis}
                    \end{scriptsize}
                \end{tikzpicture}
            \end{subfigure}
            \begin{subfigure}[b]{0.48\textwidth}
                \centering
                \begin{tikzpicture}
                    \begin{scriptsize}
                    \begin{axis}[
                        xmin = 0, xmax = 24,
                        ymin = 0, ymax = 1.05,
                        width = \textwidth,
                        height = \textwidth,
                        xlabel = $t$,
                        xtick = {0, 4, 8, 12, 16, 20, 24},
                        ]
                        \addplot[color=blue,]
                        coordinates {(0,1) (1,0.382) (2,0.282) (3,0.237) (4,0.214) (5,0.202) (6,0.199) (7,0.207) (8,0.232) (9,0.18) (10,0.155) (11,0.14) (12,0.132) (13,0.127) (14,0.126) (15,0.129) (16,0.138) (17,0.118) (18,0.107) (19,0.1) (20,0.095) (21,0.093) (22,0.093) (23,0.094) (24,0.099)};
                        \addlegendentry{$\rho = 0.45$}
                        \addplot[color=red,]
                        coordinates {(0,1) (1,0.283) (2,0.229) (3,0.205) (4,0.189) (5,0.178) (6,0.172) (7,0.175) (8,0.232) (9,0.155) (10,0.137) (11,0.129) (12,0.122) (13,0.118) (14,0.115) (15,0.116) (16,0.138) (17,0.107) (18,0.098) (19,0.094) (20,0.09) (21,0.088) (22,0.086) (23,0.087) (24,0.099)};
                        \addlegendentry{$\rho = 0.2$}
                        \addplot[color=green,]
                        coordinates {(0,1) (1,0.252) (2,0.224) (3,0.204) (4,0.189) (5,0.178) (6,0.169) (7,0.163) (8,0.232) (9,0.145) (10,0.136) (11,0.128) (12,0.122) (13,0.117) (14,0.114) (15,0.111) (16,0.138) (17,0.102) (18,0.098) (19,0.094) (20,0.09) (21,0.088) (22,0.085) (23,0.084) (24,0.099)};
                        \addlegendentry{$\rho = 0$}
                    \end{axis}
                    \end{scriptsize}
                \end{tikzpicture}
            \end{subfigure}
            \caption{Distances (left) and covariances (right) for the graph in Figure \ref{fig:LayerGraphExample2} between the points $P_0$ and $P_t$ for $\alpha = 10$ and $\beta = 0.3$. Covariances have been generated via the Dagum kernel with parameters $\alpha=1$, $\beta = 2$ and $\xi = 0.5$ (see Table \ref{tab:complMonotExamples}).}
            \label{fig:exampleDists6_CircTimeDists2}
        \end{figure}
        
        The distances and covariances in Figure \ref{fig:exampleDists6_CircTimeDists} show the effect of the parameter $\beta$ on our construction. It adds a linear component $\beta^2 t$, which allows to calibrate the distance (and, as a result, the covariance) between the points at different time instants. In such a way, the effect of the periodicity is increased by setting a low $\beta$ and becomes negligible when $\beta$ grows. Furthermore, notice the spikes the covariance functions show: they perfectly embody the periodic setting of a process, as introduced in Subsection \ref{ssec:introCircular}. It is in order to remark that, although isotropic covariances are decreasing functions of the spatial distances, Figures \ref{fig:exampleDists6_CircTimeDists} and \ref{fig:exampleDists6_CircTimeDists2} show valid covariance functions, as the distance of our setting is completely different from the Euclidean distance on $\R^n$ as it takes into account the spatio-temporal structure of the time-evolving graph.\par
                
        In Figure \ref{fig:exampleDists6_CircTimeDists2}, it is possible to visualise the effect of the partial correlation parameter $\rho\in [0,\frac{1}{2})$. First, notice that its role is particularly significant when the weight $\alpha$ is high. Indeed, for low $\alpha$'s, the covariance structure of the vertices given by the inverse laplacian matrix is dominant. Yet, when $\alpha$ is high, the nodes at different time instants are considered close to each other and the resulting distances given by the sheer first line of (\ref{eq:covZ_Periodic}) are low. Thus, the parameter $\rho$ (which enters in the second line of (\ref{eq:covZ_Periodic})) has a greater influence. Clearly, the greater is $\rho$, the lower is the distance, as it correlates different Brownian bridge realisations.

\section{Conclusion: impact of this research} 
    \label{sec:conclusion}
    
    The research presented in this paper provides the fundamentals to a part of literature that had considered networks evolution to a very limited extent. The impact of this research is multifarious:
    \begin{enumerate}
        \item temporal evolution of networks is preoccupying many scientists in both theoretical and applied disciplines. In network design problems (NDP), the optimal selection of subgraphs is a major issue. Several applications of generalised NDP arise in the fields of telecommunications, transportation and biology, and the reader is referred to \cite{feremans2003generalized}, with the references therein. Dealing with these kind of problems under the framework presented here will be a major task. 
        \item Most of the approaches about generalised networks are computer based, and the attention has been largely put on algorithmic complexity rather than on mathematical and statistical accuracy, see \cite{harrington1976generalized} and \cite{glover1978generalized}, for instance. Here, we have provided an analytic effort, and the computational burden has been taken care thanks to the assumption of temporal Markovianity. 
        \item Several emerging fields are increasingly working over generalised networks. Yet, temporal evolution has been considered under overly restrictive assumptions. Crime data on networks have been considered with strong emphasis on computer and algorithmic complexity \citep{chen2004crime}. More recently there has been a considerable interest in the statistical assessment of such data over networks, and the reader is referred to \cite{bernasco2010statistical}, and the more recent paper by \cite{muschert}, with the references therein. The connections (correlations and causalities) studied through generalised networks are of main interests in psychometrics \citep{epskamp2017generalized} and in genetics \citep{yip2007gene}, and the literature is starting to produce contributions where the interconnections are continuously defined over the edges and not only at the nodes.
    \end{enumerate}
    Several fundamental research questions are then inspired on the contribution in this paper. To mention a few, regularity properties of temporally evolving stochastic processes defined over generalised networks can now be studied thanks to the framework provided in this paper. Another field of research that will benefit from our research is that of stream flows, where the graphs are oriented (think about a river with currents in a given direction). In finance, there is a fertile literature on causality, with special emphasis on graph causality, and the reader is referred to the monumental effort in \cite{lopez2022causal}. Studying causality under continuity---as in our approach---is definitely a major challenge for the future. \par
    Other major theoretical challenges involve the definition of multivariate processes over generalised networks. Further, it is imperative to relax the assumptions of stationarity and isotropy. All these will be major challenges for future researches. 

\section*{Acknowledgements}
    The authors are grateful to Havard Rue for insightful discussions during the preparation of this manuscript. This project has also been sustained by the prompt help of Valeria Simoncini.

\clearpage

\appendix

\section{Mathematical background}
    \label{app:mathBackground}
	The material throughout is standard and there are many classical references at hand. We mention \cite{lauritzen1996graphical} amongst others. 
    \begin{definition}[Simple, connected graph]
        \label{def:simpleConnGraph}
        Let $V$ a finite set (called \emph{vertices} or \emph{nodes}) and let $E\subseteq V\times V$ a set of connections (called \emph{edges}). Then the pair $G:=(V,E)$ is called a \emph{graph} or a \emph{network}. We said that a graph $G$ is \emph{simple} if it is \emph{undirected} (videlicet, whenever $(v_1,v_2)\in E$, then necessarily $(v_2,v_1)\in E$) and if it has no self-loops (namely, for all $v \in V$, $(v,v)\notin E$). For an undirected graph $G$, two nodes $v_1,v_2 \in V$ are \emph{adjacent} if $(v_1,v_2)\in E$. We write $v_1 \sim v_2$ if $v_1$ and $v_2$ are adjacent, $v_1 \not \sim v_2$ if they are not. \par
        A \emph{path} between two nodes $v\neq w \in V$ is a finite sequence of vertices $(v=v_1, v_2, \dots, v_p = w)$ such that, $\forall i \in \curly{1, \dots, p-1}$, $(v_i,v_{i+1})\in E$. A \emph{connected component} of a graph $G$ is a maximal subset of vertices $V'\subseteq V$ such that, for each pair of nodes $v,w \in V'$, there is a path between $v$ and $w$. A graph $G$ is \emph{connected} if there is only one connected component, namely if there is a path between each pair of nodes. 
    \end{definition}

    \begin{definition}[Moore-Penrose generalised inverse]
        \label{def:MoorePenroseInverse}
        Let $M \in \R^{n \times n}$ a matrix. Then its \emph{Moore-Penrose generalised inverse} is the unique matrix $M^+ \in \R^{n \times n}$ such that:
        \begin{itemize}
            \item $M M^+ M = M$ and $M^+ M M^+ = M^+$,
            \item $M M^+ = (M M^+)^\top$ and $M^+ M = (M^+ M)^\top$.
        \end{itemize}
    \end{definition}

    \begin{definition}[$g$-embedding, isometric embedding]
        Let $(X, d)$ be a semi-metric space and $g:D_X^d \to [0,+\infty)$ a function, where $D_X^d$ denotes the diameter of $X$. Then $(X,d)$ is said to have a $g$-embedding \citep{anderes2020}
        into a Hilbert space $(H,\norm{\cdot}_H)$, written $(X,d) \overset{g}{\hookrightarrow} H$, if there exists a map $\psi : X \to H$ such that, for all $x,y\in X$,
        \begin{equation*}
            g(d(x,y))=\norm{\psi(x) - \psi(y)}_H.
        \end{equation*}
        If $g$ is the identity map, then it is called \emph{isometric embedding}.
    \end{definition}

\section{Definition of isotropic kernels on arbitrary domains}
    \label{app:isotropicKernelsGeneral}
    In this brief Section, we state and enrich some crucial results of \cite{anderes2020} that can be used in a variety of different frameworks. While Theorem \ref{theo:covarianceConstruction} furnishes a straightforward recipe for the definition of kernels as compositions of variograms and completely monotone functions, Proposition \ref{prop:variogramDistanceProperties} characterises the separation property and the triangle inequality for a variogram. As a sheer application of the former, we obtain the proof of Proposition \ref{prop:kernel_comp}.

    \begin{theorem}
        \label{theo:covarianceConstruction}
        Let $Z$ be a stochastic process defined on a set $X$ such that $\Exp{Z^2(x)}<+\infty$ for all $x\in X$. Define
        \begin{align}
            \label{eq:distanceAsVariogramGeneral}
            d:X\times X & \to \R_0^+\nonumber\\
            (x_1,x_2) & \mapsto \gamma_Z(x_1,x_2)=\Var\round{Z(x_1)-Z(x_2)}.
        \end{align}
        In addition, let $\psi$ be a non-constant completely monotone function on $[0,+\infty)$. Then the following holds:
        \begin{enumerate}
            \item $(X,d)\overset{\sqrt{\cdot}}{\hookrightarrow} H$ for some Hilbert space $H$, that is: there exist a Hilbert space $H$ and a function $\xi:X\to H$ such that, given $x_1,x_2\in X$:
            \begin{equation*}
                \sqrt{d(x_1,x_2)}=\norm{\xi(x_1)-\xi(x_2)}_H;
            \end{equation*}
            \item the function $(x_1,x_2)\mapsto \psi(d(x_1,x_2))$ is positive semidefinite;
            \item if, in addition, $d$ is a semi-metric on $X$ (see Proposition \ref{prop:variogramDistanceProperties} for a useful characterisation), then $(x_1,x_2)\mapsto \psi(d(x_1,x_2))$ is strictly positive definite.
        \end{enumerate}
    \end{theorem}

    \begin{proposition}
        \label{prop:variogramDistanceProperties}
        Let $Z$, $X$ and $d$ as in Theorem \ref{theo:covarianceConstruction}. Then:
        \begin{enumerate}
            \item $(X,d)$ is a semi-metric space if and only if, for all $x_1,x_2\in X$, $Z(x_1)=Z(x_2)$ almost surely implies $x_1=x_2$;
            \item $d$ satisfies the triangular inequality if and only if, for all $x_1,x_2,x_3\in X$, it holds:
            \begin{equation*}
                \Cov{Z(x_1)-Z(x_2)}{Z(x_3)-Z(x_2)}\geq 0.
            \end{equation*}
        \end{enumerate}
    \end{proposition}

\section{Mathematical proofs}
    \label{app:proofs}
    
    \begin{proof}[Proof of Proposition \ref{prop:expressionFor_k_Z}]
      
    \noindent From the definition of $Z_V$, we have that
            \begin{align}
                \label{eq:covZ_V}
                \Cov{Z_V(u_1)}{Z_V(u_2)}&=\round{1-\delta_1}\round{1-\delta_2}L^+[\ul u_1,\ul u_2] + \round{1-\delta_1}\delta_2 L^+[\ul u_1,\ol u_2]\nonumber \\
                &\quad +\delta_1 \round{1-\delta_2}L^+[\ol u_1,\ul u_2] + \delta_1 \delta_2 L^+[\ol u_1, \ol u_2]\nonumber \\
                &=\boldsymbol{\delta}_1^\top \, L^+\square{(\ul u_1, \ol u_1),(\ul u_2, \ol u_2)}\,\boldsymbol{\delta}_2.
            \end{align}\par
            Let us now consider the process $Z_E$. For the sake of simplicity, here we set $e_1:=(\ul u_1,\ol u_1)$ and $e_2:=(\ul u_2,\ol u_2)$. By construction, the covariance between $Z_E(u_1)$ and $Z_E(u_2)$ is null whenever $\lf(e_2)\ne \lf(e_1)$. Furthermore, 
            \[
                \Cov{Z_E(u_1)}{Z_E(u_2)}=\ell(e_1)\round{\min(\delta_1,\delta_2)-\delta_1\delta_2}
            \] 
            if $e_1=e_2 \in E_T$ and
            \[
                \Cov{Z_E(u_1)}{Z_E(u_2)}=\sqrt{\ell(e_1) \ell(e_2)}\,k_T\round{\abs{t(\ul u_1)-t(\ul u_2)}}\round{\min\round{\delta_1, \delta_2}-\delta_1 \delta_2}
            \]
            if $e_1,e_2\in E_S$ and $\lf(e_1)=\lf(e_2)$.
            By noticing that the covariance expression for former case is actually a special case of the one of the latter (recall that $k_T(0)=1$), we can summarise the covariance function of $Z_E$ for each couple of points $u_1,u_2\in \boldsymbol{G}$ as follows:
            \begin{equation}
                \label{eq:covZ_E}
                k_{Z_E}(u_1, u_2)=\mathbbm{1}_{\lf(e_1)=\lf(e_2)}\sqrt{\ell(e_1) \ell(e_2)}\,k_T\round{\abs{t(\ul u_1)-t(\ul u_2)}}\round{\min\round{\delta_1, \delta_2}-\delta_1 \delta_2}.
            \end{equation}
            Notice that the process $Z_E$ is defined on all the vertices $V$ (it is zero) and that the expression (\ref{eq:covZ_E}) is meaningful even when any of the points $u_1$ and $u_2$ belongs to $V$. Indeed, if, say, $u_1\in V$, then $\delta_1 \in \curly{0,1}$ regardless of which incident edge $(u,v)$ is taken in the expression $u_1=(u, v, \delta)$. As a consequence, the last factor in (\ref{eq:covZ_E}) vanishes and the covariance is therefore null. Finally, notice that, since $Z_V$ and $Z_E$ are independent, the covariance function of $Z$ is simply the sum of (\ref{eq:covZ_V}) and (\ref{eq:covZ_E}).
        \end{proof}
        
    \begin{proof}[Proof of Proposition \ref{prop:semiMetric}]
      
    \noindent Symmetry, non-negativeness and the implication $u_1=u_2 \Longrightarrow d(u_1,u_2)=0$ follow immediately from (\ref{eq:distAsVariogram}). Therefore, we just need to show that $d(u_1,u_2)=0 \Longrightarrow u_1=u_2$. From (\ref{eq:distAsVariogram}), if $d(u_1,u_2)=0$, then $Z(u_1)=Z(u_2)$ almost surely. As a consequence:
            \[
                Z_V(u_1)-Z_V(u_2)=-Z_E(u_1)+Z_E(u_2).
            \]
            Being $Z_V$ and $Z_E$ independent, necessarily $Z_V(u_1)-Z_V(u_2)=0$ and $-Z_E(u_1)+Z_E(u_2)=0$, hence $Z_V(u_1)=Z_V(u_2)$ and $Z_E(u_1)=Z_E(u_2)$ a.s.. Now $Z_V(u_1)$ and $Z_V(u_2)$ are linear combinations of $Z_V(V)$, that is: $Z_V(u_1)=x_1^\top Z_V(V)$ and $Z_V(u_2)=x_2^\top Z_V(V)$ for some $x_1,x_2\in \R^N$, where $N=\abs V$. More specifically, $x_1$ and $x_2$ have the following structure (here we assume that the vertices $V$ are ordered, so that $\ul u_i$ comes before $\ol u_i$, for $i\in\curly{1,2}$):
            \begin{equation*}
                \begin{cases}
                    x_1^\top=\begin{bmatrix}
                        \boldsymbol 0^\top & 1-\delta_e(u_1) & \boldsymbol 0^\top & \delta_e(u_1) & \boldsymbol 0^\top
                    \end{bmatrix}\\
                    x_2^\top=\begin{bmatrix}
                        \boldsymbol 0^\top & 1-\delta_e(u_2) & \boldsymbol 0^\top & \delta_e(u_2) & \boldsymbol 0^\top
                    \end{bmatrix},
                \end{cases}
            \end{equation*}
            where the $\boldsymbol 0$'s represent vectors of zeroes of the appropriate length (possibly zero). Since $Z_V(u_1)=Z_V(u_2)$ a.s., it follows that $(x_1-x_2)^\top Z_V(V)=0$ a.s., that is: $x_1-x_2=\lambda \boldsymbol{1}_{N}$ for some $\lambda\in \R$. Hence, 
            \begin{equation*}
                \lambda=\lambda \frac{\boldsymbol{1}_{N}^\top \boldsymbol{1}_{N}}{N}=\frac{1}{N}\boldsymbol{1}_{N}^\top (x_1-x_2)=\frac{1}{N}\round{\boldsymbol{1}_{N}^\top x_1 - \boldsymbol{1}_{N}^\top x_2} = 0.
            \end{equation*}
            This means that $x_1=x_2$, that is $u_1=u_2$.
        \end{proof}
        
    \begin{proof}[Proof of Proposition \ref{prop:notAMetric}]
      
    \noindent Consider the equivalent simple graph represented in Figure \ref{fig:triangleIneqCounterExample}, where all the weights of are $1$, exception made for the edges $(A_0,B_0)$ and $A_2,B_2$, which have weights $\varepsilon$ and $\frac{1}{\varepsilon}$ respectively, for a sufficiently small $\varepsilon>0$. Considering the vertices in the order $A_0,B_0,A_1,\dots,B_2$, the laplacian matrix $L$ is
            \begin{footnotesize}
                \begin{equation*}
                    L=\begin{bmatrix}
                        1+\varepsilon   &   -\varepsilon  &   -1  &   0   &   0  &   0\\
                        -\varepsilon  &   1+\varepsilon   &   0   &   -1  &   0  &   0\\
                        -1  &   0   &   3   &   -1  &   -1  &   0\\
                        0   &   -1  &   -1  &   3   &   0   &   -1\\
                        0   &   0   &   -1  &   0   &   1+\frac{1}{\varepsilon}   &   -\frac{1}{\varepsilon}\\
                        0   &   0   &   0   &   -1  &   -\frac{1}{\varepsilon}  &   1+\frac{1}{\varepsilon}
                    \end{bmatrix}.
                \end{equation*}
            \end{footnotesize}
            Let now consider the points $P:=\round{0,A_0,B_0,\frac{1}{2}}$, $Q:=\round{1,A_1,B_1,\frac{1}{2}}$ and $R:=\round{2,A_2,B_2,\frac{1}{2}}$. We will show that, for any $\gamma>0$, for $\varepsilon$ sufficiently small, it holds 
            \begin{equation}
                \label{eq:triangleIneqNegated}
                d(P,Q)+d(Q,R)<d(P,R).
            \end{equation}
            First, let us rewrite and simplify a bit (\ref{eq:triangleIneqNegated}). Notice that here all the $\delta$'s are $\frac{1}{2}$.
            \begin{align}
                (\ref{eq:triangleIneqNegated}) &\Longleftrightarrow k_Z(P,P)+k_Z(Q,Q)-2k_Z(P,Q)\nonumber\\
                &\quad+k_Z(Q,Q)+k_Z(R,R)-2k_Z(Q,R)\nonumber\\
                &\quad<k_Z(P,P)+k_Z(R,R)-2k_Z(P,R)\nonumber\\
                &\Longleftrightarrow k_Z(Q,Q)-k_Z(P,Q)-k_Z(Q,R)<-k_Z(P,R)\nonumber\\
                &\Longleftrightarrow \frac{1}{4} \boldsymbol{1}_{2}^\top \, L^+\square{(A_1,B_1),(A_1, B_1)}\,\boldsymbol{1}_{2} + \sqrt{1 \cdot 1}\,\gamma^{0} \cdot \frac{1}{4} \nonumber\\
                &\quad-\frac{1}{4} \boldsymbol{1}_{2}^\top \, L^+\square{(A_0,B_0),(A_1, B_1)}\,\boldsymbol{1}_{2} - \sqrt{\frac{1}{\varepsilon} \cdot 1}\,\gamma^{1} \cdot \frac{1}{4} \nonumber\\
                &\quad-\frac{1}{4} \boldsymbol{1}_{2}^\top \, L^+\square{(A_1,B_1),(A_2, B_2)}\,\boldsymbol{1}_{2} - \sqrt{1 \cdot \varepsilon}\,\gamma^{1} \cdot \frac{1}{4}\nonumber\\
                &\quad < -\frac{1}{4} \boldsymbol{1}_{2}^\top \, L^+\square{(A_0,B_0),(A_2, B_2)}\,\boldsymbol{1}_{2} - \sqrt{\frac{1}{\varepsilon} \cdot \varepsilon}\,\gamma^{2} \cdot \frac{1}{4}\nonumber\\
                &\Longleftrightarrow \boldsymbol{1}_{2}^\top \round{L^+\square{(A_1,B_1),(A_1, B_1)}+L^+\square{(A_0,B_0),(A_2, B_2)}} \boldsymbol{1}_{2}\nonumber\\
                &\quad +\boldsymbol{1}_{2}^\top \round{-L^+\square{(A_0,B_0),(A_1, B_1)}-L^+\square{(A_1,B_1),(A_2, B_2)}} \boldsymbol{1}_{2}\nonumber\\
                &\quad < -1+\frac{\gamma}{\sqrt\varepsilon}+\gamma\sqrt{\varepsilon}-\gamma^2.
                \label{eq:inequalityExampleTriangularFails}
            \end{align}
            Notice that the right-hand size of the last inequality (\ref{eq:inequalityExampleTriangularFails}) is not limited for any $\gamma>0$ when $\varepsilon\to0^+$. As a consequence, it is sufficient to show that the left-hand size is limited when $\varepsilon\to 0^+$. Indeed the left-hand side of the last inequality is a sheer signed sum of 16 elements of the matrix $L^+$. This sum is surely not greater than 
            \begin{align*}
                16\max_{i,j\in\curly{1,...,6}}\abs{L^+[i,j]}\leq 16\max_{i\in\curly{1,...,6}}\abs{L^+[i,i]}.
            \end{align*}
            Now, in our case, the main diagonal of $L^+$ is given by:
            \begin{scriptsize}
                \begin{align*}
                    \diag(L^+)=\bigg[
                        &\frac{10 \varepsilon^2+39 \varepsilon+34}{36 \left(\varepsilon^2+3 \varepsilon+1\right)},\frac{10 \varepsilon^2+39 \varepsilon+34}{36 \left(\varepsilon^2+3 \varepsilon+1\right)},\frac{10 \varepsilon^2+27 \varepsilon+10}{36 \left(\varepsilon^2+3 \varepsilon+1\right)},\frac{10 \varepsilon^2+27 \varepsilon+10}{36 \left(\varepsilon^2+3 \varepsilon+1\right)},\\
                        &\frac{34 \varepsilon^2+39 \varepsilon+10}{36 \left(\varepsilon^2+3 \varepsilon+1\right)},\frac{34 \varepsilon^2+39 \varepsilon+10}{36 \left(\varepsilon^2+3 \varepsilon+1\right)}
                    \bigg].
                \end{align*}
            \end{scriptsize}
            As all the entries are continuous functions of $\varepsilon\in[0,1]$, they are limited. Since the maximum of (a finite number of) limited functions on the same domain is limited, the left-hand of (\ref{eq:inequalityExampleTriangularFails}) is limited as well. This concludes the proof.
        \end{proof}
        
    \begin{proof}[Proof of Proposition \ref{prop:semiMetricPeriodic}]
      
    \noindent  The proof is very similar to the one of Proposition \ref{prop:semiMetric}. Also in this case, we get symmetry, non-negativeness and the implication $u_1=u_2 \Longrightarrow d(u_1,u_2)=0$ immediately from (\ref{eq:distAsVariogram}). Let us show that $d(u_1,u_2)=0 \Longrightarrow u_1=u_2$. From (\ref{eq:distAsVariogram}), if $d(u_1,u_2)=0$, then $Z(u_1)=Z(u_2)$ almost surely. As a consequence:
            \[
                Z_V(u_1)-Z_V(u_2)=-Z_E(u_1)+Z_E(u_2)-\beta W(t_1)+\beta W(t_2).
            \]
            Since $Z_V$ is independent from $Z_E$ and $W$, it must be $Z_V(u_1)=Z_V(u_2)$ a.s.. Following the same argument of the proof of Proposition \ref{prop:semiMetric}, we obtain $(\ul u_1, \ol u_1, \delta_1)=(\ul u_2, \ol u_2, \delta_2)$. It remains to be shown that $t_1=t_2$. Again, using $Z(u_1)=Z(u_2)$, and the independence between $W_t$ and both $Z_V$ and $Z_E$, we obtain $\beta W(t_1)=\beta W(t_2)$ a.s., that is $t_1=t_2$.
        \end{proof}
        
    \begin{proof}[Proof of Theorem \ref{theo:covarianceConstruction}]
            
            \begin{enumerate}
                \item Define $\Tilde{d}:=\sqrt{d}$. Since $d=\Tilde{d}^2$ is a variogram, it is conditionally negative semidefinite: as a consequence, by \citet[Theorem 6]{anderes2020}, $(X,\Tilde d) \overset{id}{\hookrightarrow} H$ for some Hilbert space $H$. Therefore $(X,d) \overset{\sqrt{\cdot}}{\hookrightarrow} H$.
                \item Follows immediately from the previous point and \citet[Corollary 1]{anderes2020}.
                \item If $Z(x_1)=Z(x_2)$ almost surely implies $x_1=x_2$, then $(X,d)$ is a semi-metric space by Proposition \ref{prop:variogramDistanceProperties}. Thus, by \citet[Corollary 1]{anderes2020}, $(x_1,x_2)\mapsto C(d(x_1,x_2))$ is strictly positive definite.
            \end{enumerate}
        \end{proof}
    
        \begin{proof}[Proof of Proposition \ref{prop:variogramDistanceProperties}]
            \begin{enumerate}
                \item $(X,d)$ is a semi-metric space iff $d(x_1,x_2)=0$ implies $x_1=x_2$. But $d(x_1,x_2)=0$ is equivalent to $Z(x_1)=Z(x_2)$ almost surely.
                \item Without loss of generality, we can restrict the proof to a zero-mean process $X$. Indeed, both the variance and the covariance do not change if we change the mean of their arguments. The proof consists in the following chain of equivalences. Let $x_1,x_2,x_3 \in X$.
                \begin{align*}
                    & d(x_1,x_2)+d(x_2,x_3)\geq d(x_1,x_3)  \\
                    \Longleftrightarrow & \Var\round{Z(x_1)-Z(x_2)}+\Var\round{Z(x_2)-Z(x_3)}\geq \Var\round{Z(x_1)-Z(x_3)} \\
                     \Longleftrightarrow &\Var Z(x_1)+\Var Z(x_2) - 2\, \Cov{Z(x_1)}{Z(x_2)} \\
                    & +\Var Z(x_2) + \Var Z(x_3) - 2\,\Cov{Z(x_2)}{Z(x_3)} \geq \\
                    & \Var Z(x_1) + \Var Z(x_3) - 2\,\Cov{Z(x_1)}{Z(x_3)}  \\
                    \Longleftrightarrow & \Var Z(x_2) - \Cov{Z(x_1)}{Z(x_2)} - \Cov{Z(x_2)}{Z(x_3)} \geq \\
                    & - \Cov{Z(x_1)}{Z(x_3)}  \\
                     \Longleftrightarrow & \,\Exp{Z^2(x_2)+Z(x_1)Z(x_3)-Z(x_1)Z(x_2)-Z(x_2)Z(x_3)}\geq 0 \\
                     \Longleftrightarrow & \,\Exp{(Z(x_2)-Z(x_1))(Z(x_2)-Z(x_3))}\geq 0 \\
                     \Longleftrightarrow & \,\Cov{Z(x_2)-Z(x_1)}{Z(x_2)-Z(x_3)} \geq 0
                \end{align*}
            \end{enumerate}
        \end{proof}

\end{document}